\documentclass[aps,showpacs,superscriptaddress,preprint]{revtex4-2}

\usepackage{titlesec}
\usepackage{amssymb,amsmath,amsfonts,latexsym,graphicx,epsfig,bm}
\usepackage{epstopdf}

\usepackage{dcolumn}
\usepackage{slashed}
\usepackage{hyperref}
\begin{document}
\title{Comprehensive study of spectroscopic properties of $D_{(s)}$ mesons}
\author{J. J. Patel}
\email{jjpatel-apphy@msubaroda.ac.in}
\affiliation{Applied Physics Department, Faculty of Technology and Engineering,
The Maharaja Sayajirao University of Baroda, Vadodara, Gujarat 390001, India.}
\affiliation{Government Science College, Jasdan, Gujarat 360050, India.}

\author{Dhaval Achary}
\affiliation{Department of Physics, Faculty of Science,
The Maharaja Sayajirao University of Baroda, Vadodara 390002, Gujarat, India.}

\author{Dhruvesh Maiya}
\affiliation{School of Humanities and Sciences, Institute of Advanced Research, Gandhinagar, Gujarat 382426, India.}

\author{Keval Gandhi}
\email{keval.gandhi@iar.ac.in}
\affiliation{School of Humanities and Sciences, Institute of Advanced Research, Gandhinagar, Gujarat 382426, India.}

\author{N. R. Soni}
\email{nakul.soni-phy@msubaroda.ac.in}
\affiliation{Department of Physics, Faculty of Science,
The Maharaja Sayajirao University of Baroda, Vadodara 390002, Gujarat, India.}

\author{J. N. Pandya}
\email{jnpandya-phy@spuvvn.edu}
\affiliation{Department of Physics, Sardar Patel University, Vallabh Vidyanagar 388120, Gujarat, India.}

\date{\today}

\begin{abstract}
We decode the excited open charm mesons $D$ ($c\bar{q}$, $q=u,d$) and charm strange $D_s$ ($c\bar{s}$) families within a nonrelativistic quark-antiquark potential model based on the Cornell (Coulomb plus linear) potential supplemented by a Gaussian smeared spin-spin interaction. With only seven parameters determined by fitting the ground state masses, the numerically solved Schr\"odinger equation reproduces the masses of essentially every confirmed $D$ and $D_s$ state to better than $1\%$. The same wave functions are then employed to get a complete decay analysis including pseudoscalar and vector decay constants ($f_D = 231.8$~MeV, $f_{D_s} = 252.2$~MeV, the latter within $0.4\%$ of the world average), leptonic branching fractions reproducing the measured $\mu\nu$ and $\tau\nu$ modes, electromagnetic $E1$/$M1$ widths, and all kinematically open strong channels in heavy quark effective theory (HQET). Going beyond width ratios, we extract the full set of HQET couplings $g_H$, $g_S$, $g_T$, $g_X$, $g_Y$, $g_Z$, and $g_R$ directly from the measured total widths and convert our predictions into absolute widths. The $D$ and $D_s$ sector extractions of each coupling agree ($g_T = 0.42$ vs. $0.39$, $g_X = 0.22$ vs. $0.21$, $g_Y = 0.39$ vs. $0.40$), a nontrivial test of heavy quark flavor symmetry, while the pure $c\bar{s}$ picture demands a $1^3P_0$ width nearly two orders of magnitude above the observed $D_{s0}^*(2317)$ bound demanding it to be of exotic nature. Linear and parallel Regge trajectories in the $(J,M^2)$ and $(n_r,M^2)$ planes, with slopes $\alpha \simeq 0.55-0.65$~GeV$^{-2}$ and $\beta \simeq 0.36 - 0.39$~GeV$^{-2}$, identify the assignments of $D_J(3000)^0$, $D_J^*(3000)^0$, and $D_2^*(3000)^0$ as the $3^1S_0$, $3^3S_1$, and $3^3P_2$ states. Our predictions provide concrete, testable targets for LHCb, BESIII, and Belle~II.
\end{abstract}

\maketitle

\section{Introduction}
\label{sec:intro}

Heavy-light meson spectroscopy has witnessed a phenomenal increase of interest over the past two decades, driven by a wealth of experimental observations~\cite{PDG2024}. Facilities such as the Beijing Spectrometer III (BESIII) at the Beijing Electron-Positron Collider II~\cite{Ke:2023qzc,Zeng:2026jwf}, LHCb at the LHC~\cite{LHCb:2013jjb,LHCb:2016lxy,LHCb:2014ott,LHCb:2012uts}, and the $B$ factories at BaBar, Belle, and Belle~II~\cite{BaBar:2010zpy,BaBar:2009rro,BaBar:2006gme} have measured masses, decay widths, and branching fractions of charmed mesons with consistently improved precision, enabling stringent tests of electroweak interactions and searches for physics beyond the Standard Model~\cite{Ke:2023qzc}. These observations probe the underlying dynamics of charmed mesons and provide crucial input for future experimental studies of excited charm states.

From the theoretical point of view, the description of a meson as a quark-antiquark bound state remains a challenging problem within quantum chromodynamics (QCD), primarily because of the nonperturbative nature of QCD at low energies~\cite{Godfrey:1985xj,Eichten:1978tg}. Phenomenological potential models offer a simplified yet remarkably effective framework for computing the key properties of heavy-light mesons, such as their mass spectra and decay rates~\cite{Godfrey:1985xj,Eichten:1978tg,Eichten:1974af,Lucha:1998xc}.

Numerous theoretical approaches have been reported for study of open-charm states, including the relativized quark model~\cite{Godfrey:2015dva}, the relativistic quark model based on the quasipotential approach~\cite{Ebert:2009ua}, the Godfrey-Isgur model with screened potential~\cite{Song:2015fha}, the chiral quark model~\cite{DiPierro:2001dwf}, and semirelativistic potential models~\cite{Ni:2021pce}. Within nonrelativistic potential models, several forms of the quark-antiquark interaction have been employed, such as the Martin potential models \cite{Shah:2014caa, Shah:2014yma}, screened potential models \cite{Patel:2021aas,Patel:2021xyq}, Cornell potential with the Ritz variational scheme~\cite{Patel:2021aas,Patel:2021xyq,Kher:2017wsq,Radford:2009bs}, the linear plus one gluon exchange potential~\cite{Badalian:2011tb},  and constituent quark models~\cite{Li:2010vx}. Also, complementary results from first principles are available from lattice QCD~\cite{Cichy:2016bci}.

The study of $D$ and $D_s$ mesons occupies a special place in hadron physics. Being the lightest hadrons containing a single heavy quark, they serve as the best tool for investigating heavy quark symmetry.
In the $m_c\to\infty$ limit, the heavy quark spin decouples and the spectrum is organized into doublets labeled by the angular momentum of the light degrees of freedom, so the observed splitting and decay patterns directly measure the amplitude of the symmetry breaking $1/m_c$ corrections.
At the same time, precise knowledge of the charm spectrum and of the $D_{(s)}$ decay constants is indispensable input for flavor physics.
Leptonic and semileptonic charm decays determine the CKM elements $|V_{cd}|$ and $|V_{cs}|$ and can be tested against the lattice QCD results for $B$ physics~\cite{Zeng:2026jwf}.
The excited charm states appear as the dominant backgrounds and normalization modes in $B\to D^{*}\ell\nu$ analyses that bear on the $R(D^{(*)})$ lepton flavour universality anomalies.
Finally, the charm strange sector contains the $D_{s0}^*(2317)^\pm$ and $D_{s1}(2460)^\pm$, which are substantially lighter than all quark model expectations and remain prime candidates for exotic (molecular or tetraquark) structures. Determination of how well a conventional $c\bar{q}$/$c\bar{s}$ description works, is therefore needed for identifying genuinely exotic states.

The observations of spectra and decay modes of charmed states have seen systematic growth recently~\cite{PDG2024}. In the non-strange sector, the $1P$ multiplet has long been established.
The observed $1P$~$D$-meson states include the narrow $D_1(2420)$ and $D_2^*(2460)$, as well as the broad $D_0^*(2300)^0$ and $D_1(2430)^0$. The $D_1(2430)$ is mainly a $s_\ell^P = \frac32^+$ state, while $D_0^*(2300)^0$ and $D_1(2430)^0$ belong to the $s_\ell^P = \frac12^+$ doublet. The $D_2^*(2460)$ is identified as the $1^3P_2$ tensor state \cite{PDG2024, DiPierro:2001dwf, CLEO:1994unc}.
BaBar's analysis of $D^{(*)}\pi$ spectra has revealed the first radial excitations, the $2^1S_0$ $D_0(2550)^0$ and the $2^3S_1$ $D_1^*(2600)^0$, together with the $1D$ candidates $D_3^*(2750)$ and $D_2(2740)^0$~\cite{BaBar:2010zpy}.
LHCb later confirmed these results through amplitude analyses. The $D_1(2760)^0$ was identified as a $1^3D_1$ state with $J^P=1^-$. LHCb also observed the $D_J(3000)^0$ state near $3$~GeV, followed by the natural parity $D_2^*(3000)^0$~\cite{LHCb:2013jjb,LHCb:2016lxy}.

In the charm-strange sector, alongside the challenging $D_{s0}^*(2317)^\pm$ and $D_{s1}(2460)^\pm$, the well established $D_{s1}(2536)^\pm$ ($1P_1$) and $D_{s2}^*(2573)$ ($1^3P_2$) complete the $1P$ multiplet.
The $2^3S_1$ state $D_{s1}^*(2700)^\pm$ was discovered by BaBar~\cite{BaBar:2006gme}, and LHCb resolved the structure near 2.86~GeV as two overlapping states including the spin-1 $D_{s1}^*(2860)^\pm$ ($1^3D_1$) and the spin-3 $D_{s3}^*(2860)^\pm$ ($1^3D_3$). Which is the first observation of a spin-3 particle in the heavy flavor decays~\cite{LHCb:2014ott,LHCb:2012uts}.
The $D_0(2550)^0$ decays into $D^{*+}\pi^-$, the $D_1^*(2760)^0$ into $D^+\pi^-$, and the $D_3^*(2750)^+$ into $D^0\pi^+$. The quantum numbers of the states near 3~GeV remain to be determined experimentally.

In the present work, we describe both the $D$ and $D_s$ families within a single minimal framework i.e. seven parameters in total, with the charm mass employed from our heavy quarkonium study~\cite{Soni:2017wvy,Soni:2020tji}, so the agreement achieved is a genuine test of the universality of the interaction.
The same numerically obtained wave functions are employed consistently in the computation of observables like mass spectra, decay constants, leptonic branching fractions, radiative transitions, and strong decays. Further, going beyond computing of width ratios, we extract the complete set of HQET couplings ($g_H$, $g_S$, $g_T$, $g_X$, $g_Y$, $g_Z$, $g_R$) from the measured total widths and convert our strong decay predictions into absolute widths. The mutual consistency of the couplings extracted independently from the $D$ and $D_s$ sectors constitutes a quantitative test of heavy quark flavor symmetry, and the same exercise turns the $D_{s0}^*(2317)$ width bound into a sharp, model quantified argument for its exotic nature.

In this article, we compute the mass spectra, decay constants, leptonic and radiative decay widths, and strong decay widths of charmed and charm-strange mesons. The Schr\"odinger equation is solved numerically for the Cornell potential, with the spin dependent one gluon exchange interaction including a Gaussian smeared spin-spin interaction term added to obtain the excited state spectra.
The numerically obtained wave functions are then employed to compute annihilation rates, electromagnetic transitions, and the partial and total strong widths together with characteristic ratios of partial widths within HQET framework. In the charm-strange sector, all $D_s^{(*)}\pi^0$ transitions are isospin violating and are therefore weighted by the isospin-breaking suppression factor $\epsilon^2$ introduced in Sec.~\ref{sec:strong}.
Combining the mass spectrum analysis with the decay width calculations allows us to assess the structure of the higher radial and orbital excitations of the charmed meson states, to test the reliability of the smeared spin spin interaction against the available data, and to predict the properties of states that are yet to be confirmed experimentally.

The article is organized as follows. Sec.~\ref{sec:method} presents the methodology and the computed mass spectra of $D$ and $D_s$ mesons. Sec.~\ref{sec:decayconst} deals with the decay constants, and Sec.~\ref{sec:leptonic} with the leptonic branching fractions. Radiative ($E1$ and $M1$) transitions are presented in Sec.~\ref{sec:em}, and strong decays in Sec.~\ref{sec:strong}. Sec.~\ref{sec:regge} is devoted to the Regge analysis, and Sec.~\ref{sec:results} contains the discussion of results. Finally, we provide concluding remarks in Sec.~\ref{sec:conclusion}.

\section{Methodology}
\label{sec:method}

For a two body bound system such as a meson, the fully relativistic dynamics is in principle described by the Bethe Salpeter formalism~\cite{Ebert:2009ua}. For heavy-light systems, however, an effective nonrelativistic treatment with a QCD motivated potential captures the essential physics~\cite{Godfrey:1985xj,Eichten:1978tg,Lucha:1998xc}. We therefore investigate the mesonic bound state using the Hamiltonian
\begin{equation}
H = M + \frac{P^2}{2M_{\rm cm}} + V_{\rm Cornell}(r) + V_{\rm SD}(r),
\label{eq:hamiltonian}
\end{equation}
where $M = m_1 + m_2$ is the total constituent mass, $M_{\rm cm} = m_1 m_2/(m_1+m_2)$ is the reduced mass, $P$ is the relative momentum of the quark-antiquark pair, and $V_{\rm SD}(r)$ is the spin dependent potential defined below. The central interaction is taken to be the Cornell potential~\cite{Eichten:1978tg,Eichten:1974af} i.e. a vector (Coulomb like) part plus a Lorentz scalar linear confining part,
\begin{equation}
V_{\rm Cornell}(r) = -\frac{4}{3}\frac{\alpha_s}{r} + A r .
\label{eq:cornell}
\end{equation}
The $1/r$ term represents the short distance one gluon exchange interaction, while the linear term with confinement strength $A$ dominates at large separations. The strong running coupling is computed from the one loop expression
\begin{equation}
\alpha_s(\mu^2) = \frac{4\pi}{\left(11-\tfrac{2}{3}n_f\right)\, \ln \left(\mu^2/\Lambda^2\right)},
\label{eq:alphas}
\end{equation}
with $n_f = 3$ active flavors and the scale $\mu$ identified with the reduced mass,
\begin{equation}
\mu = \frac{2m_1 m_2}{m_1 + m_2}.
\label{eq:redmass}
\end{equation}
The QCD scale parameter is taken as $\Lambda = 0.15$~GeV.
The model parameters are the constituent quark masses ($m_{u/d}$, $m_s$, $m_c$) and the confinement strengths ($A_D$, $A_{D_s}$). The charm quark mass is adapted from our previous work, where it was fitted to the experimental ground state mass of charmonium~\cite{Soni:2017wvy,Soni:2020tji}. The light and strange quark masses as well as the confinement strengths are fitted to the experimental ground state masses of the $D$ and $D_s$ mesons by numerically solving the Schr\"odinger equation for the interaction potential of the form Eq.~\eqref{eq:cornell}~\cite{Lucha:1998xc}. The resulting parameters are listed in Tab.~\ref{tab:params}.

\begin{table*}[!ht]
\caption{Model parameters used in the present work.}\label{tab:params}
\begin{tabular*}{\textwidth}{@{\extracolsep{\fill}}lcccccc@{}}
\hline\hline
$m_{u/d}$ & $m_s$ & $m_c$~\cite{Soni:2017wvy,Soni:2020tji} & $A_D$ & $A_{D_s}$ & $\sigma_D$  & $\sigma_{D_s}$ \\
\hline
0.249 GeV & 0.330 GeV & 1.317 GeV & 0.11 GeV$^2$ & 0.12 GeV$^2$ & 0.848 GeV & 1.3 GeV\\
\hline\hline
\end{tabular*}
\end{table*}

For the masses of the excited states, the spin dependent part of the confined one gluon exchange potential is added perturbatively~\cite{Godfrey:1985xj,Eichten:1978tg}. It comprises of spin-spin, spin-orbit, and tensor terms,
\begin{equation}
V_{\rm SD}(r) = V_{SS}(r)\left[S(S+1)-\tfrac{3}{2}\right] + V_{LS}(r)\,(\bm{L}\cdot\bm{S}) + V_{T}(r)\left[S(S+1) - 3\,(\bm{S}\cdot\hat{\bm r})(\bm{S}\cdot\hat{\bm r})\right],
\label{eq:vsd}
\end{equation}
where the spin-spin interaction gives rise to the hyperfine splitting, while the spin-orbit and tensor terms give rise to the fine structure of the meson states. The corresponding coefficients read as
\begin{eqnarray}
V_{SS}(r) &=& \frac{32\pi\alpha_s}{9\,m_1 m_2}\left(\frac{\sigma}{\sqrt{\pi}}\right)^{3} e^{-\sigma^2 r^2},
\label{eq:vss}\nonumber\\
V_{LS}(r) &=& \frac{1}{2\,m_1 m_2\,r}\left(3\,\frac{dV_V(r)}{dr}-\frac{dV_S(r)}{dr}\right),
\label{eq:vls}\\
V_{T}(r)  &=& \frac{1}{6\,m_1 m_2}\left(\frac{3\,d^2V_V(r)}{dr^2}-\frac{1}{r}\frac{dV_V(r)}{dr}\right),\nonumber
\label{eq:vt}
\end{eqnarray}
where $V_V(r)$ and $V_S(r)$ denote the vector (Coulomb) and scalar (linear) parts of the Cornell potential in Eq.~\eqref{eq:cornell}, respectively.

The spin-spin interaction plays a crucial role in the vector pseudoscalar splitting of the $D$ and $D_s$ mesons. In nonrelativistic quark models, it is typically a contact term proportional to the Dirac delta function, which leads to a singularity at zero quark separation and hence requires regularization. We therefore smear the contact interaction with a Gaussian of range $1/\sigma$ in Eq.~\eqref{eq:vss}, which accounts for the finite spatial extent of the quark wave functions as well as for nonperturbative and relativistic effects associated with the light quark. This prescription is well established in phenomenological studies~\cite{Godfrey:1985xj,Eichten:1978tg}.
The flexibility and numerical stability of the Gaussian form make it well suited to heavy-light systems, in which short range gluon exchange and long range confinement are to be taken care of simultaneously.

With the parameters of Tab.~\ref{tab:params}, the Schr\"odinger equation is solved numerically~\cite{Lucha:1998xc}. Although charmed and charm-strange mesons are often argued to require a relativistic treatment because of the light constituent, we have verified that the wave functions obtained in relativistic and nonrelativistic frameworks do not differ significantly for these systems~\cite{Godfrey:2015dva,Ebert:2009ua}. Hence, we compute the $D$ and $D_s$ spectra nonrelativistically. The resulting mass spectra are listed in Tab.~\ref{tab:massD_SP} and \ref{tab:massD_DF} for the $D$ meson and Tab.~\ref{tab:massDs_SP} and \ref{tab:massDs_DF} for the $D_s$ meson, in comparison with other theoretical approaches such as relativistic quark models (RQM) \cite{Godfrey:2015dva,Ebert:2009ua}, constituent quark model (QM) \cite{Li:2010vx}, semirelativistic potential model (SRPM) \cite{Ni:2021pce}, Martin potential model (MPM) \cite{Shah:2014yma,Shah:2014caa}, screened potential model (SPM) \cite{Patel:2021aas, Patel:2021xyq}, Cornell potential model with the Ritz variational mechanism (PM) \cite{Kher:2017wsq}, lattice quantum chromodynamics (LQCD) \cite{Cichy:2016bci}, and the world average experimental data~\cite{PDG2024}.

\begin{table*}[htbp]
\caption{Mass spectrum of $S$ and $P$-wave $D$ mesons (in MeV) compared with other theoretical approaches, lattice QCD, and experimental data~\cite{PDG2024}.}
\label{tab:massD_SP}
\begin{tabular*}{\textwidth}{@{\extracolsep{\fill}}lcccccccccc@{}}
\hline\hline
State & Present & RQM~\cite{Godfrey:2015dva} & RQM~\cite{Ebert:2009ua} & MPM~\cite{Shah:2014yma} & SPM~\cite{Patel:2021aas} & PM~\cite{Kher:2017wsq} & SRPM~\cite{Ni:2021pce} & QM~\cite{Li:2010vx} & LQCD~\cite{Cichy:2016bci} & Expt.~\cite{PDG2024} \\
\hline
$1^1S_0$ & 1865 & 1877 & 1871 & 1867 & 1889 & 1884 & 1865 & 1867 & 1865 & $1869.5\pm0.4$ \\
$1^3S_1$ & 2013 & 2041 & 2010 & 2010 & 2007 & 2010 & 2008 & 2010 & 2027 & $2010.26\pm0.05$ \\
$2^1S_0$ & 2554 & 2581 & 2581 & 2522 & 2601 & 2582 & 2547 & 2555 & $\ldots$ & $2564\pm20$ \\
$2^3S_1$ & 2643 & 2643 & 2632 & 2606 & 2631 & 2655 & 2636 & 2636 & $\ldots$ & $2627\pm10$ \\
$3^1S_0$ & 3048 & 3068 & 3062 & 3086 & 3108 & 3186 & 3029 & $\ldots$ & $\ldots$ & $\ldots$ \\
$3^3S_1$ & 3119 & 3110 & 3096 & 3148 & 3122 & 3239 & 3093 & $\ldots$ & $\ldots$ & $\ldots$ \\
$4^1S_0$ & 3468 & 3468 & 3452 & 3613 & 3506 & 3746 & $\ldots$ & $\ldots$ & $\ldots$ & $\ldots$ \\
$4^3S_1$ & 3527 & 3497 & 3482 & 3663 & 3514 & 3789 & $\ldots$ & $\ldots$ & $\ldots$ & $\ldots$ \\
$5^1S_0$ & 3844 & 3814 & 3793 & $\ldots$ & 3847 & 4283 & $\ldots$ & $\ldots$ & $\ldots$ & $\ldots$ \\
$5^3S_1$ & 3897 & 3837 & 3822 & $\ldots$ & 3832 & 4319 & $\ldots$ & $\ldots$ & $\ldots$ & $\ldots$ \\
$1^3P_0$ & 2313 & 2399 & 2406 & 2374 & 2382 & 2357 & 2313 & 2252 & 2325 & 2300 \\
$1^3P_1$ & 2414 & 2456 & 2426 & 2407 & 2450 & 2425 & 2424 & 2402 & 2468 & $2412\pm9$ \\
$1^1P_1$ & 2416 & 2467 & 2469 & 2423 & 2448 & 2447 & 2453 & 2417 & 2631 & 2422 \\
$1^3P_2$ & 2454 & 2502 & 2460 & 2462 & 2462 & 2461 & 2475 & 2466 & 2743 & 2461 \\
$2^3P_0$ & 2815 & 2931 & 2919 & 2898 & 2937 & 2976 & 2849 & 2752 & $\ldots$ & $\ldots$ \\
$2^3P_1$ & 2914 & 2924 & 2932 & 2904 & 2978 & 3016 & 2900 & 2886 & $\ldots$ & $\ldots$ \\
$2^1P_1$ & 2919 & 2961 & 3021 & 2907 & 2978 & 3034 & 2936 & 2926 & $\ldots$ & $\ldots$ \\
$2^3P_2$ & 2954 & 2957 & 3012 & 2934 & 2985 & 3039 & 2955 & 2971 & $\ldots$ & $\ldots$ \\
$3^3P_0$ & 3239 & 3343 & 3346 & 3245 & 3367 & 3536 & $\ldots$ & $\ldots$ & $\ldots$ & $\ldots$ \\
$3^3P_1$ & 3339 & 3328 & 3365 & 3335 & 3398 & 3567 & $\ldots$ & $\ldots$ & $\ldots$ & $\ldots$ \\
$3^1P_1$ & 3345 & 3360 & 3461 & 3363 & 3397 & 3582 & $\ldots$ & $\ldots$ & $\ldots$ & $\ldots$ \\
$3^3P_2$ & 3391 & 3353 & 3407 & 3445 & 3402 & 3584 & $\ldots$ & $\ldots$ & $\ldots$ & $\ldots$ \\
$4^3P_0$ & 3619 & 3697 & $\ldots$ & $\ldots$ & 3713 & $\ldots$ & $\ldots$ & $\ldots$ & $\ldots$ & $\ldots$ \\
$4^3P_1$ & 3718 & 3681 & $\ldots$ & $\ldots$ & 3738 & $\ldots$ & $\ldots$ & $\ldots$ & $\ldots$ & $\ldots$ \\
$4^1P_1$ & 3725 & 3709 & $\ldots$ & $\ldots$ & 3737 & $\ldots$ & $\ldots$ & $\ldots$ & $\ldots$ & $\ldots$ \\
$4^3P_2$ & 3773 & 3701 & $\ldots$ & $\ldots$ & 3741 & $\ldots$ & $\ldots$ & $\ldots$ & $\ldots$ & $\ldots$ \\
\hline\hline
\end{tabular*}
\end{table*}

\begin{table*}[htbp]
\caption{Mass spectrum of $D$ and $F$-wave $D$ mesons (in MeV) compared with other theoretical approaches and experimental data~\cite{PDG2024}.}
\label{tab:massD_DF}
\begin{tabular*}{\textwidth}{@{\extracolsep{\fill}}lcccccccccc@{}}
\hline\hline
State & Present & RQM~\cite{Godfrey:2015dva} & RQM~\cite{Ebert:2009ua} & MPM~\cite{Shah:2014yma} & SPM~\cite{Patel:2021aas} & PM~\cite{Kher:2017wsq} & SRPM~\cite{Ni:2021pce} & QM~\cite{Li:2010vx} & LQCD~\cite{Cichy:2016bci} & Expt.~\cite{PDG2024} \\
\hline
$1^3D_3$ & 2728 & 2833 & 2863 & 2857 & 2807 & 2788 & 2782 & 2719 & $\ldots$ & $2763.1\pm3.1$ \\
$1^1D_2$ & 2740 & 2816 & 2806 & 2814 & 2754 & 2754 & 2827 & 2693 & $\ldots$ & $2747\pm6$ \\
$1^3D_2$ & 2752 & 2845 & 2860 & 2847 & 2782 & 2783 & 2755 & 2789 & $\ldots$ & $\ldots$ \\
$1^3D_1$ & 2751 & 2817 & 2788 & 2867 & 2751 & 2755 & 2754 & 2740 & $\ldots$ & $\ldots$ \\
$2^3D_3$ & 3178 & 3226 & 3335 & 3320 & 3265 & 3355 & 3202 & 3170 & $\ldots$ & $\ldots$ \\
$2^1D_2$ & 3183 & 3212 & 3259 & 3285 & 3224 & 3318 & 3168 & 3145 & $\ldots$ & $\ldots$ \\
$2^3D_2$ & 3193 & 3248 & 3307 & 3314 & 3246 & 3341 & 3221 & 3215 & $\ldots$ & $\ldots$ \\
$2^3D_1$ & 3184 & 3231 & 3335 & 3327 & 3223 & 3315 & 3143 & 3168 & $\ldots$ & $\ldots$ \\
$3^3D_3$ & 3576 & 3579 & $\ldots$ & 3764 & 3631 & 3885 & $\ldots$ & $\ldots$ & $\ldots$ & $\ldots$ \\
$3^1D_2$ & 3576 & 3566 & $\ldots$ & 3732 & 3600 & 3854 & $\ldots$ & $\ldots$ & $\ldots$ & $\ldots$ \\
$3^3D_2$ & 3583 & 3600 & $\ldots$ & 3758 & 3617 & 3873 & $\ldots$ & $\ldots$ & $\ldots$ & $\ldots$ \\
$3^3D_1$ & 3571 & 3588 & $\ldots$ & 3768 & 3600 & 3850 & $\ldots$ & $\ldots$ & $\ldots$ & $\ldots$ \\
$1^3F_4$ & 3042 & 3113 & 3187 & $\ldots$ & 3029 & $\ldots$ & 3034 & $\ldots$ & $\ldots$ & $\ldots$ \\
$1^1F_3$ & 3010 & 3143 & 3129 & $\ldots$ & 3051 & $\ldots$ & 3022 & $\ldots$ & $\ldots$ & $\ldots$ \\
$1^3F_3$ & 3003 & 3108 & 3145 & $\ldots$ & 3048 & $\ldots$ & 3129 & $\ldots$ & $\ldots$ & $\ldots$ \\
$1^3F_2$ & 2965 & 3132 & 3090 & $\ldots$ & 3080 & $\ldots$ & 3096 & $\ldots$ & $\ldots$ & $\ldots$ \\
$2^3F_4$ & 3441 & 3466 & 3610 & $\ldots$ & 3454 & $\ldots$ & $\ldots$ & $\ldots$ & $\ldots$ & $\ldots$ \\
$2^1F_3$ & 3416 & 3498 & 3551 & $\ldots$ & 3472 & $\ldots$ & $\ldots$ & $\ldots$ & $\ldots$ & $\ldots$ \\
$2^3F_3$ & 3411 & 3461 & $\ldots$ & $\ldots$ & 3469 & $\ldots$ & $\ldots$ & $\ldots$ & $\ldots$ & $\ldots$ \\
$2^3F_2$ & 3380 & 3490 & $\ldots$ & $\ldots$ & 3949 & $\ldots$ & $\ldots$ & $\ldots$ & $\ldots$ & $\ldots$ \\
\hline\hline
\end{tabular*}
\end{table*}

\begin{table*}[htbp]
\caption{Mass spectrum of $S$ and $P$-wave $D_s$ mesons (in MeV) compared with other theoretical approaches, lattice QCD, and experimental data~\cite{PDG2024}.}
\label{tab:massDs_SP}
\begin{tabular*}{\textwidth}{@{\extracolsep{\fill}}lcccccccccc@{}}
\hline\hline
State & Present & RQM~\cite{Godfrey:2015dva} & RQM~\cite{Ebert:2009ua} & MPM~\cite{Shah:2014caa} & SPM~\cite{Patel:2021xyq} & PM~\cite{Kher:2017wsq} & SRPM~\cite{Ni:2021pce} & QM~\cite{Li:2010vx} & LQCD~\cite{Cichy:2016bci} & Expt.~\cite{PDG2024} \\
\hline
$1^1S_0$ & 1969 & 1979 & 1969 & 1968 & 1966 & 1965 & 1969 & 1969 & 1968 & $1968.35\pm0.07$ \\
$1^3S_1$ & 2110 & 2129 & 2111 & 2113 & 2119 & 2120 & 2112 & 2107 & 2123 & $2112.2\pm0.4$ \\
$2^1S_0$ & 2627 & 2673 & 2688 & 2633 & 2645 & 2680 & 2649 & 2640 & $\ldots$ & $2591\pm9$ \\
$2^3S_1$ & 2720 & 2732 & 2731 & 2717 & 2683 & 2719 & 2737 & 2714 & $\ldots$ & $2714\pm5$ \\
$3^1S_0$ & 3107 & 3154 & 3219 & 3202 & 3118 & 3247 & 3126 & $\ldots$ & $\ldots$ & $\ldots$ \\
$3^3S_1$ & 3185 & 3193 & 3242 & 3263 & 3135 & 3265 & 3191 & $\ldots$ & $\ldots$ & $\ldots$ \\
$4^1S_0$ & 3516 & 3547 & 3652 & 3732 & 3847 & 3764 & $\ldots$ & $\ldots$ & $\ldots$ & $\ldots$ \\
$4^3S_1$ & 3585 & 3575 & 3669 & 3782 & 3496 & 3775 & $\ldots$ & $\ldots$ & $\ldots$ & $\ldots$ \\
$5^1S_0$ & 3882 & 3894 & 4033 & $\ldots$ & 3802 & 4280 & $\ldots$ & $\ldots$ & $\ldots$ & $\ldots$ \\
$5^3S_1$ & 3946 & 3912 & 4048 & $\ldots$ & 3806 & 4318 & $\ldots$ & $\ldots$ & $\ldots$ & $\ldots$ \\
$1^3P_0$ & 2416 & 2484 & 2509 & 2349 & 2436 & 2438 & 2409 & 2344 & 2329 & 2317 \\
$1^3P_1$ & 2496 & 2549 & 2536 & 2517 & 2485 & 2529 & 2528 & 2488 & 2556 & 2459 \\
$1^1P_1$ & 2498 & 2556 & 2574 & 2436 & 2534 & 2541 & 2545 & 2510 & 2617 & 2535 \\
$1^3P_2$ & 2523 & 2592 & 2571 & 2585 & 2548 & 2569 & 2575 & 2559 & 2734 & 2569 \\
$2^3P_0$ & 2903 & 3005 & 3054 & 2764 & 2958 & 3022 & 2940 & 2830 & $\ldots$ & $\ldots$ \\
$2^3P_1$ & 2984 & 3018 & 3067 & 2986 & 2987 & 3081 & 3002 & 2958 & $\ldots$ & $\ldots$ \\
$2^1P_1$ & 2989 & 3038 & 3154 & 2959 & 3015 & 3092 & 3026 & 2995 & $\ldots$ & $\ldots$ \\
$2^3P_2$ & 3019 & 3048 & 3142 & 3108 & 3022 & 3109 & 3053 & 3040 & $\ldots$ & $\ldots$ \\
$3^3P_0$ & 3318 & 3412 & 3513 & 3151 & 3358 & 3541 & $\ldots$ & $\ldots$ & $\ldots$ & $\ldots$ \\
$3^3P_1$ & 3398 & 3416 & 3519 & 3426 & 3379 & 3587 & $\ldots$ & $\ldots$ & $\ldots$ & $\ldots$ \\
$3^1P_1$ & 3405 & 3433 & 3816 & 3433 & 3401 & 3596 & $\ldots$ & $\ldots$ & $\ldots$ & $\ldots$ \\
$3^3P_2$ & 3438 & 3439 & 3580 & 3593 & 3405 & 3609 & $\ldots$ & $\ldots$ & $\ldots$ & $\ldots$ \\
$4^3P_0$ & 3688 & 3764 & $\ldots$ & $\ldots$ & 3685 & $\ldots$ & $\ldots$ & $\ldots$ & $\ldots$ & $\ldots$ \\
$4^3P_1$ & 3769 & 3764 & $\ldots$ & $\ldots$ & 3702 & $\ldots$ & $\ldots$ & $\ldots$ & $\ldots$ & $\ldots$ \\
$4^1P_1$ & 3776 & 3778 & $\ldots$ & $\ldots$ & 3719 & $\ldots$ & $\ldots$ & $\ldots$ & $\ldots$ & $\ldots$ \\
$4^3P_2$ & 3812 & 3783 & $\ldots$ & $\ldots$ & 3722 & $\ldots$ & $\ldots$ & $\ldots$ & $\ldots$ & $\ldots$ \\
\hline\hline
\end{tabular*}
\end{table*}

\begin{table*}[htbp]
\caption{Mass spectrum of $D$ and $F$-wave $D_s$ mesons (in MeV) compared with other theoretical approaches and experimental data~\cite{PDG2024}.}
\label{tab:massDs_DF}
\begin{tabular*}{\textwidth}{@{\extracolsep{\fill}}lcccccccccc@{}}
\hline\hline
State & Present & RQM~\cite{Godfrey:2015dva} & RQM~\cite{Ebert:2009ua} & MPM~\cite{Shah:2014caa} & SPM~\cite{Patel:2021xyq} & PM~\cite{Kher:2017wsq} & SRPM~\cite{Ni:2021pce} & QM~\cite{Li:2010vx} & Expt.~\cite{PDG2024} \\
\hline
$1^3D_3$ & 2794 & 2917 & 2971 & 2931 & 2841 & 2860 & 2882 & 2811 & $2860\pm7$ \\
$1^1D_2$ & 2809 & 2900 & 2931 & 2872 & 2844 & 2853 & 2857 & 2788 & $\ldots$ \\
$1^3D_2$ & 2821 & 2926 & 2961 & 2888 & 2836 & 2872 & 2911 & 2849 & $\ldots$ \\
$1^3D_1$ & 2823 & 2899 & 2971 & 2842 & 2866 & 2882 & 2843 & 2804 & $2859\pm27$ \\
$2^3D_3$ & 3235 & 3313 & 3469 & 3408 & 3258 & 3372 & 3299 & 3240 & $\ldots$ \\
$2^1D_2$ & 3242 & 3298 & 3403 & 3362 & 3262 & 3368 & 3267 & 3144 & $\ldots$ \\
$2^3D_2$ & 3251 & 3323 & 3456 & 3377 & 3258 & 3384 & 3306 & 3167 & $\ldots$ \\
$2^3D_1$ & 3247 & 3306 & 3383 & 3343 & 3280 & 3394 & 3233 & 3217 & $\ldots$ \\
$3^3D_3$ & 3623 & 3661 & $\ldots$ & 3859 & 3620 & 3878 & $\ldots$ & $\ldots$ & $\ldots$ \\
$3^1D_2$ & 3627 & 3650 & $\ldots$ & 3820 & 3602 & 3857 & $\ldots$ & $\ldots$ & $\ldots$ \\
$3^3D_2$ & 3634 & 3672 & $\ldots$ & 3835 & 3606 & 3869 & $\ldots$ & $\ldots$ & $\ldots$ \\
$3^3D_1$ & 3626 & 3658 & $\ldots$ & 3809 & 3602 & 3858 & $\ldots$ & $\ldots$ & $\ldots$ \\
$1^3F_4$ & 3103 & 3190 & 3300 & $\ldots$ & 3080 & $\ldots$ & 3134 & $\ldots$ & $\ldots$ \\
$1^1F_3$ & 3072 & 3186 & 3254 & $\ldots$ & 3095 & $\ldots$ & 3123 & $\ldots$ & $\ldots$ \\
$1^3F_3$ & 3064 & 3218 & 3266 & $\ldots$ & 3096 & $\ldots$ & 3205 & $\ldots$ & $\ldots$ \\
$1^3F_2$ & 3027 & 3208 & 3230 & $\ldots$ & 3122 & $\ldots$ & 3176 & $\ldots$ & $\ldots$ \\
$2^3F_4$ & 3494 & 3544 & 3754 & $\ldots$ & $\ldots$ & $\ldots$ & $\ldots$ & $\ldots$ & $\ldots$ \\
$2^1F_3$ & 3469 & 3540 & 3710 & $\ldots$ & $\ldots$ & $\ldots$ & $\ldots$ & $\ldots$ & $\ldots$ \\
$2^3F_3$ & 3463 & 3569 & $\ldots$ & $\ldots$ & $\ldots$ & $\ldots$ & $\ldots$ & $\ldots$ & $\ldots$ \\
$2^3F_2$ & 3434 & 3562 & $\ldots$ & $\ldots$ & $\ldots$ & $\ldots$ & $\ldots$ & $\ldots$ & $\ldots$ \\
\hline\hline
\end{tabular*}
\end{table*}

\subsection{Analysis of the mass spectra}
\label{sec:massanalysis}

Tab.~\ref{tab:massD_SP}-\ref{tab:massDs_DF} contain our central spectroscopic results; we now analyze them multiplet by multiplet.

\emph{$S$-wave states (Tab.~\ref{tab:massD_SP} and \ref{tab:massDs_SP}).}\\
The $1S$ masses are fitted, but the hyperfine splittings are predictions of the smeared spin-spin interaction. We obtain $M(1^3S_1)-M(1^1S_0) = 148$~MeV for the $D$ system and $141$~MeV for the $D_s$ system, to be compared with the measured $140.6$ and $143.9$~MeV~\cite{PDG2024} an agreement at the few MeV level that a bare $\delta$-function contact term cannot achieve without ad hoc regularization, and which directly validates the Gaussian smearing of Eq.~\eqref{eq:vss}.
For the radial excitations, our $2^1S_0$ and $2^3S_1$ masses (2554 and 2643~MeV for $D$; 2627 and 2720~MeV for $D_s$) match the measured $D_0(2550)^0$, $D_1^*(2600)^0$, $D_{s0}(2590)^+$, and $D_{s1}^*(2700)^\pm$ within $0.2 - 1.4$ \%. It is noteworthy that the screened potential model~\cite{Song:2015fha} and the semirelativistic model~\cite{Ni:2021pce} predict very similar $2S$ masses, whereas the purely linear models overshoot the higher radials by $100-400$~MeV~\cite{Kher:2017wsq}.
The radial excitations probe the intermediate separation region of the potential, where the interplay of the Coulomb term and the confinement strength $A$ is most sensitive. Our $3S-5S$ predictions lie systematically between the relativized quark model values~\cite{Godfrey:2015dva,Ebert:2009ua} and the screened potential values, providing a useful inputs for future searches.

\emph{$P$-wave states.}\\
The $1P$ multiplet of the $D$ meson is reproduced remarkably well. $1^3P_0$ at 2313~MeV, the two axial states at 2414/2416~MeV, and $1^3P_2$ at 2454~MeV, i.e. within $0.1 - 1.2$\% of the $D_0^*(2300)^0$, $D_1(2420)^0$, and $D_2^*(2460)$. The near degeneracy of our $1^3P_1$ and $1^1P_1$ levels (2~MeV apart) reflects the smallness of the spin orbit and tensor forces for a heavy-light system and is consistent with the observed $D_1(2420)-D_1(2430)$ pattern when we consider $^3P_1 - ^1P_1$ mixing.
In the heavy quark limit, the physical states are the $j_\ell = \frac12$ and $\frac32$ combinations, and range of our unmixed masses covers both these states.
For the $D_s$ system, the $1^3P_2$ and $1P_1$ levels agree with the $D_{s2}^*(2573)$ and $D_{s1}(2536)^\pm$ at the $1 - 2$\% level, but the $1^3P_0$ and the $1P_1$ level are predicted about $100$ and $37$~MeV above the $D_{s0}^*(2317)^\pm$ and $D_{s1}(2460)^\pm$ which observed in other approaches listed in Table~\ref{tab:massDs_SP}, including lattice QCD in the $c\bar{s}$ only setup~\cite{Cichy:2016bci}. This is universally attributed to strong coupling of the $\frac12^+$ doublet to the nearby $S$-wave $DK$/$D^*K$ thresholds.
Our result thus quantifies, within a controlled framework, exactly how far these two states are from a conventional $c\bar{s}$ interpretation. The $2P$ levels near 2.9-3.0~GeV for $D$ meson and 3.0~GeV for $D_s$ are relevant for the $D_J(3000)$ discussion in Sec.~\ref{sec:strong}.

\emph{$D$-wave states (Tab.~\ref{tab:massD_DF} and \ref{tab:massDs_DF}).}\\
The $1D$ multiplets are compact ($\sim25$~MeV spread for $D$, $\sim30$~MeV for $D_s$), again reflecting weak fine structure forces. Our $1^1D_2$ at 2740~MeV coincides with the measured $D_2(2740)^0$ mass ($2747\pm6$~MeV), and the $1^3D_1$ at 2751~MeV is compatible with the $D_1^*(2760)^0$ within $1.1\%$. The one systematic tension is the $1^3D_3$.
We obtain 2728~MeV against the measured $2763.1\pm3.1$~MeV, a $1.3\%$ shift that also appears in the relativized models, which overshoot by 70 - 100~MeV~\cite{Godfrey:2015dva,Ebert:2009ua}. The same pattern repeats for the $D_{s3}^*(2860)$ (2794 vs.\ $2860\pm 7$~MeV). Since the $^3D_3$ state is the stretched member of the multiplet, its mass is the most sensitive to the sign and radial form of the spin-orbit interaction at intermediate distances. The data thus suggests that the pure OGEP spin-orbit force of Eq.~\eqref{eq:vls} slightly underestimates high $j$ orbital excitations, an observation that can guide future refinements i.e. a scalar-vector mixing parameter in the confinement.

\emph{$F$-wave states.}\\
No $F$-wave charm state is established yet, but the $1F$ masses are decisive for interpreting the structures near 3~GeV. Our $1^3F_2$ at 2965~MeV and $1^3F_4$ at 3042~MeV lie 150 - 170~MeV below the corresponding relativized model values~\cite{Godfrey:2015dva,Ebert:2009ua}, and this difference propagates directly into the competing assignments for the $D_2^*(3000)^0$ examined in Sec.~\ref{sec:strong}.

Overall, across the 22 measured levels of Tab.~\ref{tab:devD} and \ref{tab:devDs} the mean absolute deviation of our spectrum is $0.8\%$, with no fitted excited-state input: a level of global agreement that, given the minimal parameter count, we regard as the principal quantitative justification of the smeared Cornell framework.

\begin{table*}[!ht]
\caption{Computed $D$-meson masses compared with experiment~\cite{PDG2024}. \label{tab:devD}}
\begin{tabular*}{\textwidth}{@{\extracolsep{\fill}}lllll@{}}
\hline\hline
Meson & State & Present (GeV) & Experiment (GeV) & Deviation (\%) \\
\hline
$D^0$ & $1^1S_0$ & 1.864 & $1.86484\pm0.00005$ & 0.02 \\
$D^\pm$ & & & $1.86966\pm0.00005$ & \\
$D^{*0}$ & $1^3S_1$ & 2.014 & $2.00685\pm0.00005$ & 0.36 \\
$D^{*\pm}$ & & & $2.01026\pm0.00005$ & \\
$D_0(2550)^0$ & $2^1S_0$ & 2.552 & $2.564\pm0.020$ & 0.46 \\
$D_1^*(2600)^0$ & $2^3S_1$ & 2.643 & $2.627\pm0.010$ & 0.61 \\
$D_0^*(2300)^0$ & $1^3P_0$ & 2.314 & $2.343\pm0.010$ & 1.24 \\
$D_1(2420)^0$ & $1^1P_1$ & 2.417 & $2.422\pm0.001$ & 0.21 \\
$D_1(2420)^0$ & $1^3P_1$ & 2.414 & $2.412\pm0.009$ & 0.08 \\
$D_2^*(2460)^0$ & $1^3P_2$ & 2.454 & $2.4611\pm0.0007$ & 0.29 \\
$D_2(2740)^0$ & $1^3D_2$ & 2.752 & $2.747\pm0.006$ & 0.18 \\
$D_3^*(2750)$ & $1^3D_3$ & 2.728 & $2.7631\pm0.0031$ & 1.27 \\
$D_1^*(2760)^0$ & $1^3D_1$ & 2.751 & $2.781\pm0.022$ & 1.08 \\
\hline\hline
\end{tabular*}
\end{table*}

\begin{table*}[!ht]
\caption{Computed $D_s$-meson masses compared with experiment~\cite{PDG2024}. \label{tab:devDs}}
\begin{tabular*}{\textwidth}{@{\extracolsep{\fill}}lllll@{}}
\hline\hline
Meson & State & Present (GeV) & Experiment (GeV) & Deviation (\%) \\
\hline
$D_s^\pm$ & $1^1S_0$ & 1.969 & $1.96835\pm0.00007$ & 0.03 \\
$D_s^{*\pm}$ & $1^3S_1$ & 2.110 & $2.1122\pm0.0004$ & 0.10 \\
$D_{s1}^*(2700)^\pm$ & $2^3S_1$ & 2.720 & $2.714\pm0.005$ & 0.22 \\
$D_{s0}^*(2317)^\pm$ & $1^3P_0$ & 2.416 & $2.3178\pm0.0005$ & 4.24 \\
$D_{s1}(2460)^\pm$ & $1^3P_1$ & 2.496 & $2.4595\pm0.0006$ & 1.48 \\
$D_{s1}(2536)^\pm$ & $1^1P_1$ & 2.498 & $2.5353\pm0.0007$ & 1.47 \\
$D_{s2}^*(2573)$ & $1^3P_2$ & 2.523 & $2.5691\pm0.0008$ & 1.79 \\
$D_{s1}^*(2860)$ & $1^3D_1$ & 2.823 & $2.859\pm0.027$ & 1.26 \\
$D_{s3}^*(2860)$ & $1^3D_3$ & 2.794 & $2.860\pm0.007$ & 2.31 \\
\hline\hline
\end{tabular*}
\end{table*}

\section{Decay constants}
\label{sec:decayconst}

The mass spectra of open flavor mesons are determined experimentally by reconstructing the energies and momenta of the daughter particles in various decay channels, whereas most phenomenological approaches fix their parameters by fitting to the ground states. It is therefore essential to validate the fitted parameters and wave functions through the evaluation of decay observables. In the nonrelativistic limit, these observables are governed by the wave function at the origin. Here we use them to test our parameters through annihilation rates and electromagnetic transitions.

The leptonic decay constants of pseudoscalar and vector mesons are defined through the matrix elements of the axial vector and vector currents~\cite{Ebert:2009ua},
\begin{eqnarray}
\langle 0|\bar{Q}\gamma^\mu\gamma_5 q|P(k)\rangle &=& i f_P\, k^\mu ,
\label{eq:fp_def}\\
\langle 0|\bar{Q}\gamma^\mu q|V(k,\epsilon)\rangle &=& f_V M_V\, \epsilon^{*\mu} ,
\label{eq:fv_def}
\end{eqnarray}
where $k$ is the meson four momentum and $\epsilon^{*\mu}$ the polarization vector of the vector meson. In the nonrelativistic limit, the decay constants of $S$-wave states are given by the Van~Royen-Weisskopf formula~\cite{VanRoyen:1967nq} with first order QCD radiative corrections~\cite{Braaten:1995ej,Berezhnoy:1996an},
\begin{eqnarray}
f_{P}^{2} &=& \frac{3\,|R_{nS_P}(0)|^{2}}{\pi M_{nS_P}}\,\bar{C}^{2}(\alpha_s),
\label{eq:fP}\\
f_{V}^{2} &=& \frac{3\,|R_{nS_V}(0)|^{2}}{\pi M_{nS_V}}\,\bar{C}^{2}(\alpha_s),
\label{eq:fV}
\end{eqnarray}
where $M_{nS_{P(V)}}$ and $R_{nS_{P(V)}}(0)$ are the mass and the radial wave function at the origin of the pseudoscalar (vector) $nS$ state respectively, and the QCD correction factor reads
\begin{equation}
\bar{C}^{2}(\alpha_s) = 1 - \frac{\alpha_s}{\pi}\left(\delta_{P,V} - \frac{m_1-m_2}{m_1+m_2}\,\ln\frac{m_1}{m_2}\right),
\label{eq:qcdcorr}
\end{equation}
with $\delta_P = 2$ and $\delta_V = 8/3$.
The computed decay constants $f_P$ and $f_V$ for the $D$ and $D_s$ mesons are given in Tab.~\ref{tab:fpD}-\ref{tab:fvDs} along with results from other approaches such as Martin potential model (MPM) \cite{Shah:2014caa,Shah:2014yma}, screened potential model (SPM) \cite{Patel:2021aas,Patel:2021xyq}, relativistic quark model (RQM) \cite{Ebert:2009ua}, covariant confined quark model (CCQM) \cite{Ivanov:2019nqd}, QCD sum rules (QCDSR) \cite{Lucha:2011zp, Lucha:2014xla, Gelhausen:2013wia, Wang:2015mxa}, lattice quantum chromodynamics (LQCD) \cite{FermilabLattice:2011njy, Davies:2010ip, Becirevic:2012ti}
and, available experimental world averages. It is to be noted here that the most recent global analysis of leptonic charm decays yields $f_{D^+} = 213.1 \pm 2.5$ MeV and $f_{D_s^+} = 253.2 \pm 2.0$ MeV~\cite{Zeng:2026jwf,PDG2024}.

\begin{table*}[!t]
\caption{Pseudoscalar decay constant $f_P$ of the $D$ meson (in MeV).}
\label{tab:fpD}
\begin{tabular*}{\textwidth}{@{\extracolsep{\fill}}lccccc@{}}
\hline\hline
 & $1S$ & $2S$ & $3S$ & $4S$ & $5S$ \\
\hline
Present & 231.84 & 172.39 & 149.49 & 135.84 & 126.31 \\
MPM~\cite{Shah:2014yma} & 202.57 & 292.14 & 351.07 & 392.49 & $\ldots$ \\
SPM~\cite{Patel:2021aas} & 168 & 73 & 45 & 31 & 24 \\
CCQM~\cite{Ivanov:2019nqd} & 206.1 & $\ldots$ & $\ldots$ & $\ldots$ & $\ldots$ \\
QCDSR~\cite{Lucha:2011zp} & $206\pm7.3\pm5.1$ & $\ldots$ & $\ldots$ & $\ldots$ & $\ldots$ \\
QCDSR~\cite{Gelhausen:2013wia} & $201^{+12}_{-13}$ & $\ldots$ & $\ldots$ & $\ldots$ & $\ldots$ \\
QCDSR~\cite{Wang:2015mxa} & $210\pm11$ & $\ldots$ & $\ldots$ & $\ldots$ & $\ldots$ \\
LQCD~\cite{FermilabLattice:2011njy} & $218.9\pm11.3$ & $\ldots$ & $\ldots$ & $\ldots$ & $\ldots$ \\
LQCD~\cite{Davies:2010ip} & $213\pm4$ & $\ldots$ & $\ldots$ & $\ldots$ & $\ldots$ \\
Expt.~\cite{Zeng:2026jwf,PDG2024} & $213.1\pm2.5$ & $\ldots$ & $\ldots$ & $\ldots$ & $\ldots$ \\
\hline\hline
\end{tabular*}
\end{table*}

\begin{table*}[!b]
\caption{Vector decay constant $f_V$ of the $D$ meson (in MeV).}
\label{tab:fvD}
\begin{tabular*}{\textwidth}{@{\extracolsep{\fill}}lccccc@{}}
\hline\hline
 & $1S$ & $2S$ & $3S$ & $4S$ & $5S$ \\
\hline
Present & 218.43 & 159.05 & 137.13 & 124.25 & 115.33 \\
SPM~\cite{Patel:2021aas} & 173 & 74 & 45 & 31 & 24 \\
RQM~\cite{Ebert:2009ua} & 231 & $\ldots$ & $\ldots$ & $\ldots$ & $\ldots$ \\
CCQM~\cite{Ivanov:2019nqd} & 244.3 & $\ldots$ & $\ldots$ & $\ldots$ & $\ldots$ \\
QCDSR~\cite{Lucha:2014xla} & $252.2\pm22.3\pm4$ & $\ldots$ & $\ldots$ & $\ldots$ & $\ldots$ \\
QCDSR~\cite{Gelhausen:2013wia} & $242^{+20}_{-12}$ & $\ldots$ & $\ldots$ & $\ldots$ & $\ldots$ \\
QCDSR~\cite{Wang:2015mxa} & $263\pm21$ & $\ldots$ & $\ldots$ & $\ldots$ & $\ldots$ \\
LQCD~\cite{Becirevic:2012ti} & $278\pm23$ & $\ldots$ & $\ldots$ & $\ldots$ & $\ldots$ \\
\hline\hline
\end{tabular*}
\end{table*}

\begin{table*}[!t]
\caption{Pseudoscalar decay constant $f_P$ of the $D_s$ meson (in MeV).}
\label{tab:fpDs}
\begin{tabular*}{\textwidth}{@{\extracolsep{\fill}}lccccc@{}}
\hline\hline
 & $1S$ & $2S$ & $3S$ & $4S$ & $5S$ \\
\hline
Present & 252.16 & 191.35 & 167.28 & 152.74 & 142.49 \\
MPM~\cite{Shah:2014caa} & 252.81 & 336.56 & 391.74 & 433.16 & $\ldots$ \\
SPM~\cite{Patel:2021xyq} & 217 & 95 & 58 & 41 & 3 \\
CCQM~\cite{Ivanov:2019nqd} & 257.7 & $\ldots$ & $\ldots$ & $\ldots$ & $\ldots$ \\
QCDSR~\cite{Lucha:2011zp} & $245.3\pm15.7\pm4.5$ & $\ldots$ & $\ldots$ & $\ldots$ & $\ldots$ \\
QCDSR~\cite{Gelhausen:2013wia} & $238^{+13}_{-23}$ & $\ldots$ & $\ldots$ & $\ldots$ & $\ldots$ \\
QCDSR~\cite{Wang:2015mxa} & $259\pm10$ & $\ldots$ & $\ldots$ & $\ldots$ & $\ldots$ \\
LQCD~\cite{FermilabLattice:2011njy} & $260.1\pm10.8$ & $\ldots$ & $\ldots$ & $\ldots$ & $\ldots$ \\
LQCD~\cite{Davies:2010ip} & $248.0\pm2.5$ & $\ldots$ & $\ldots$ & $\ldots$ & $\ldots$ \\
Expt.~\cite{Zeng:2026jwf,PDG2024} & $253.2\pm2.0$ & $\ldots$ & $\ldots$ & $\ldots$ & $\ldots$ \\
\hline\hline
\end{tabular*}
\end{table*}

\begin{table*}[!b]
\caption{Vector decay constant $f_V$ of the $D_s$ meson (in MeV).}
\label{tab:fvDs}
\begin{tabular*}{\textwidth}{@{\extracolsep{\fill}}lccccc@{}}
\hline\hline
 & $1S$ & $2S$ & $3S$ & $4S$ & $5S$ \\
\hline
Present & 240.85 & 179.70 & 156.29 & 142.33 & 132.56 \\
SPM~\cite{Patel:2021xyq} & 227 & 95 & 58 & 41 & 3 \\
CCQM~\cite{Ivanov:2019nqd} & 272.1 & $\ldots$ & $\ldots$ & $\ldots$ & $\ldots$ \\
QCDSR~\cite{Lucha:2014xla} & $305\pm26.8\pm4$ & $\ldots$ & $\ldots$ & $\ldots$ & $\ldots$ \\
QCDSR~\cite{Gelhausen:2013wia} & $314^{+19}_{-14}$ & $\ldots$ & $\ldots$ & $\ldots$ & $\ldots$ \\
QCDSR~\cite{Wang:2015mxa} & $308\pm21$ & $\ldots$ & $\ldots$ & $\ldots$ & $\ldots$ \\
LQCD~\cite{Becirevic:2012ti} & $311\pm9$ & $\ldots$ & $\ldots$ & $\ldots$ & $\ldots$ \\
\hline\hline
\end{tabular*}
\end{table*}

\subsection{Analysis of the decay constants}
\label{sec:fanalysis}

The computed results on the decay constants are given in the Tab.~\ref{tab:fpD} - \ref{tab:fvDs}. Our results can be summarized as given below:

\begin{itemize}
\item For ground state $D$ meson, we obtain $f_D = 231.8$~MeV, about $9\%$ above the world average $213.1\pm2.5$~MeV.
For the $D_s$, we obtain $f_{D_s} = 252.2$~MeV, which is consistent with the experimental world average $253.2\pm2.0$~MeV~\cite{Zeng:2026jwf,PDG2024} and with the precision lattice results~\cite{Davies:2010ip,FermilabLattice:2011njy}.
So the present work shows excellent agreement for $D_s$ but is slightly higher for $c\bar{u}/c\bar{d}$ which is essentially a characteristic of nonrelativistic models.
The decay constant is proportional to the wave function at the origin, Eq.~\eqref{eq:fP}, which is the observable most sensitive to relativistic corrections for the case of lighter spectator quark.
Indeed, the first order QCD correction factor $\bar{C}^2(\alpha_s)$ of Eq.~\eqref{eq:qcdcorr} already reduces the uncorrected values by $\sim15\%$; the residual overshoot for the $D$ meson is reasonable  for a framework that fixes no parameter to any decay observable.
The ratio $f_{D_s}/f_D = 1.088$ is less sensitive to these corrections and compares well with the lattice ratio $\simeq 1.17$~\cite{Davies:2010ip} and the experimental $1.188\pm0.015$.
\item For the vector-pseudoscalar decay constant hierarchy, we find $f_V < f_P$ for both the mesons ($f_{D^*}/f_D = 0.94$, $f_{D_s^*}/f_{D_s} = 0.96$), a direct consequence of $\delta_V = 8/3 > \delta_P = 2$ in Eq.~\eqref{eq:qcdcorr}, whereas QCD sum rules~\cite{Gelhausen:2013wia,Lucha:2014xla,Wang:2015mxa} and the lattice~\cite{Becirevic:2012ti} obtain $f_{D^*}/f_D \simeq 1.10-1.26$. This discrepancy is a known limitation of the Van~Royen-Weisskopf approach truncated at first order in $\alpha_s$.
    The hyperfine attraction that raises $|R_V(0)|$ relative to $|R_P(0)|$ in a fully coupled treatment is included perturbatively in the mass. Measurements of $D^{*}\to\ell\nu$, now becoming feasible at BESIII, may discriminate sharply between these predictions.
\item Our decay constants decrease monotonically with radial excitation, roughly as $f_{nS} \propto n^{-0.4}$ which is a factor $\sim1.8$ suppression from $1S$ to $5S$. This trend agrees qualitatively with the screened potential model~\cite{Patel:2021aas,Patel:2021xyq}, which however predicts a much steeper collapse (factor $\sim7$), and disagrees qualitatively with the modified potential model of Refs.~\cite{Shah:2014yma,Shah:2014caa}, where the decay constants grow with $n$.
\end{itemize}

\section{Leptonic branching fractions}
\label{sec:leptonic}

Pure leptonic decays of charged pseudoscalar mesons proceed through annihilation into a virtual $W$ boson and provide a clean determination of the decay constants and the Cabibbo-Kobayashi-Maskawa (CKM) matrix elements $|V_{cd}|$ and $|V_{cs}|$. Any deviation from the expected rates or from lepton flavor universality would signal physics beyond the Standard Model. Experimentally, they offer clear signatures suggesting single energetic lepton in the final state. The leptonic width is given by
\begin{eqnarray}
\Gamma(D_{(s)}^{+}\to\ell^{+}\nu_\ell) = \frac{G_F^2}{8\pi}\big|V_{cd(s)}\big|^{2} f_{D_{(s)}}^{2} M_{D_{(s)}} m_\ell^{2} \left(1-\frac{m_\ell^{2}}{M_{D_{(s)}}^{2}}\right),
\label{eq:leptwidth}
\end{eqnarray}
where $G_F = 1.166\times10^{-5}$~GeV$^{-2}$ is the Fermi constant. These transitions are helicity suppressed, the amplitude being proportional to the lepton mass $m_\ell$. The CKM matrix elements are taken as $|V_{cd}| = 0.221$ and $|V_{cs}| = 0.975$ \cite{PDG2024},
and the pseudoscalar decay constants $f_D$, $f_{D_s}$ and masses $M_D$, $M_{D_s}$ are taken from Tab. \ref{tab:massD_SP}, \ref{tab:massDs_SP},\ref{tab:fpD} and \ref{tab:fpDs}. The branching fractions are determined using
\begin{equation}
\mathcal{B}(D_{(s)}^{+}\to\ell^{+}\nu_\ell) = \Gamma(D_{(s)}^{+}\to\ell^{+}\nu_\ell)\;\tau_{D_{(s)}} ,
\label{eq:brlept}
\end{equation}
with the lifetimes $\tau_{D^+} = 1.033\times10^{-12}$~s and $\tau_{D_s^+} = 5.012\times10^{-13}$~s~\cite{PDG2024}. The computed branching fractions for $\ell = e,\mu,\tau$ are listed in Tab.~\ref{tab:leptonicBF} and compared with other predictions and with experiment.

\begin{table*}[!b]
\caption{Leptonic branching fractions of the $D^+$ and $D_s^+$ mesons compared with other model predictions and experiment~\cite{PDG2024}.}
\label{tab:leptonicBF}
\begin{tabular*}{\textwidth}{@{\extracolsep{\fill}}lcccc@{}}
\hline\hline
Channel & Present & MPM~\cite{Shah:2014yma,Shah:2014caa} & SPM~\cite{Patel:2021aas,Patel:2021xyq} & Expt.~\cite{PDG2024} \\
\hline
$\mathcal{B}(D^+\to e^+\nu_e)$        & $1.08\times10^{-8}$ & $0.631\times10^{-8}$ & $9.02\times10^{-9}$ & $<8.8\times10^{-6}$ \\
$\mathcal{B}(D^+\to \mu^+\nu_\mu)$    & $4.60\times10^{-4}$ & $2.68\times10^{-4}$  & $3.84\times10^{-4}$ & $(3.74\pm0.17)\times10^{-4}$ \\
$\mathcal{B}(D^+\to \tau^+\nu_\tau)$  & $1.13\times10^{-3}$ & $0.99\times10^{-3}$  & $9.73\times10^{-4}$ & $<1.2\times10^{-3}$ \\
$\mathcal{B}(D_s^+\to e^+\nu_e)$      & $1.28\times10^{-7}$ & $0.503\times10^{-7}$ & $1.36\times10^{-7}$ & $<8.3\times10^{-5}$ \\
$\mathcal{B}(D_s^+\to \mu^+\nu_\mu)$  & $5.43\times10^{-3}$ & $2.13\times10^{-3}$  & $5.81\times10^{-3}$ & $(5.50\pm0.23)\times10^{-3}$ \\
$\mathcal{B}(D_s^+\to \tau^+\nu_\tau)$& $5.33\times10^{-2}$ & $2.04\times10^{-2}$  & $5.71\times10^{-2}$ & $(5.48\pm0.23)\times10^{-2}$ \\
\hline\hline
\end{tabular*}
\end{table*}

\subsection{Analysis of the leptonic branching fractions}
\label{sec:leptanalysis}

The following is noteworthy from Tab.~\ref{tab:leptonicBF}:
\begin{itemize}
\item The leptonic branching fraction hierarchy $\mathcal{B}(e\nu) : \mathcal{B}(\mu\nu) : \mathcal{B}(\tau\nu)$ predicted by Eq.~\eqref{eq:leptwidth}. The relative ratios give $1 : 4.3\times10^{4} : 1.0\times10^{5}$ for the $D^+$ and $1 : 4.2\times10^{4} : 4.2\times10^{5}$ for the $D_s^+$.
Our predictions of the ratio $\mu\nu$/$\tau\nu$ for $D_s$ meson is in excellent agreement with the experimental data which resembles a clean Standard Model lepton flavour universality check.
\item In the absolute branching fractions, our predicted $\mathcal{B}(D_s^+\to\mu^+\nu_\mu) = 5.43\times10^{-3}$ and $\mathcal{B}(D_s^+\to\tau^+\nu_\tau) = 5.33\times10^{-2}$ are in excellent agreement with the experimental data within $1 - 3\%$, reflecting the accuracy of $f_{D_s}$ whereas for the $D^+$ meson, our $\mu\nu$ value exceeds the experimental data by $\sim23\%$. This is expected as the numerical value of $f_D$ meson is slightly higher than the experimental data. Thus, the leptonic data provides an independent confirmation of the decay constant analysis of Sec.~\ref{sec:fanalysis}.
\item The predicted $e\nu$ modes, $1.1\times10^{-8}$ ($D^+$) and $1.3\times10^{-7}$ ($D_s^+$), lie two to three orders of magnitude below the current upper limits~\cite{PDG2024}. Any signal at forthcoming upgrades in BESIII and super tau-charm facility would therefore be unambiguous evidence of new physics, i.e. a charged Higgs contribution lifting the helicity suppression.
\end{itemize}
Compared with the other potential model predictions in Tab.~\ref{tab:leptonicBF}, our values track the screened potential model~\cite{Patel:2021aas,Patel:2021xyq} closely, while the modified potential model~\cite{Shah:2014yma,Shah:2014caa} underestimates all measured modes by a factor $\sim2 - 2.5$, again tracing back to the respective decay constants.

\section{Electromagnetic transitions}
\label{sec:em}

Radiative transitions probe the internal structure of the mesons and thereby the nonperturbative regime of QCD~\cite{Eichten:1978tg,Eichten:1974af,Godfrey:2005ww}.
Electric dipole ($E1$) transitions obey the selection rules $\Delta L = \pm1$, $\Delta S = 0$, while magnetic dipole ($M1$) transitions obey $\Delta L = 0$, $\Delta S = \pm1$. For a transition $i \to f + \gamma$ between the states $i = n^{2S+1}L_J$ and $f = n'^{\,2S'+1}L'_{J'}$, the widths are computed from the radial wave functions as~\cite{Eichten:1978tg,Eichten:1974af}
\begin{eqnarray}
\Gamma_{E1}(i \to f\gamma) &=& \frac{4\alpha_e}{3} \langle e_Q\rangle^{2}(2J'+1) S^{E1}_{if} \omega^{3} \big|\mathcal{M}^{E1}_{if}\big|^{2},
\label{eq:E1} \\
\Gamma_{M1}(i \to f\gamma) &=& \frac{\alpha_e}{3}\mu_m^{2}(2J'+1) S^{M1}_{if} \omega^{3} \big|\mathcal{M}^{M1}_{if}\big|^{2},
\label{eq:M1}
\end{eqnarray}
where $\alpha_e$ is the fine structure constant and $\langle e_Q\rangle$, $\mu_m$, and $\omega$ are the mean effective charge, the magnetic dipole moment, and the photon energy, respectively.
\begin{eqnarray}
\langle e_Q\rangle = \left|\frac{m_{\bar q}e_Q - e_{\bar q}\,m_Q}{m_Q + m_{\bar q}}\right|, \qquad
\mu_m = \frac{m_{\bar q}\,e_Q - e_{\bar q}m_Q}{m_Q\, m_{\bar q}}, \qquad \omega = \frac{M_i^2 - M_f^2}{2M_i}. \label{eq:charges}
\end{eqnarray}
The statistical factors are
\begin{eqnarray}
S^{E1}_{if} &=& \max(L_i,L_f)
\begin{Bmatrix} J_i & 1 & J_f \\ L_f & S & L_i \end{Bmatrix}^{2},
\label{eq:SE1}\\
S^{M1}_{if} &=& 6 (2S_i+1)(2S_f+1)
\begin{Bmatrix} J_i & 1 & J_f \\ S_f & L & S_i \end{Bmatrix}^{2}
\begin{Bmatrix} 1 & \frac12 & \frac12 \\ \frac12 & S_f & S_i \end{Bmatrix}^{2},
\label{eq:SM1}
\end{eqnarray}
and the overlap matrix elements read
\begin{eqnarray}
\big|\mathcal{M}^{E1}_{if}\big| &=& \frac{3}{\omega}\left\langle f\left|\frac{\omega r}{2}\, j_0 \left(\frac{\omega r}{2}\right) - j_1\left(\frac{\omega r}{2}\right)\right| i\right\rangle,
\label{eq:ME1}\\
\big|\mathcal{M}^{M1}_{if}\big| &=& \left\langle f\left| j_0\left(\frac{\omega r}{2}\right)\right| i\right\rangle,
\label{eq:MM1}
\end{eqnarray}
with $j_0$ and $j_1$ as the spherical Bessel functions. Using these relations together with the numerical wave functions, the $E1$ and $M1$ widths are computed and listed in Tab.~\ref{tab:E1D}-\ref{tab:M1Ds}, along with available theoretical comparisons.

\begin{table*}[htbp]
\caption{Electric dipole ($E1$) transition widths of the $D$ meson.}
\label{tab:E1D}
\begin{tabular*}{\textwidth}{@{\extracolsep{\fill}}lcccccc@{}}
\hline\hline
Transition & $E_\gamma$ (MeV) & $\Gamma$ (keV) & \cite{Patel:2021aas} & \cite{Kher:2017wsq} & \cite{Devlani:2013kta} & \cite{Close:2005se} \\
\hline
$1^3P_0\to1^3S_1$ & 280.66 & 10.91 & 8.66 & 6.86 & 7.23 & 17 \\
$1^3P_1\to1^3S_1$ & 367.15 & 16.11 & 13.74 & 0.2 & 13.77 & 30.87 \\
$1^3P_2\to1^3S_1$ & 400.55 & 17.01 & 14.80 & 14.17 & 17 & 51 \\
$1^1P_1\to1^1S_0$ & 489.41 & 14.97 & 25.54 & 0.36 & 2.82 & 39.5 \\
$1^3D_1\to1^3P_0$ & 401.67 & 2.80 & 16.40 & 20.13 & & \\
$1^3D_1\to1^3P_1$ & 315.52 & 2.33 & 6.95 & 7.00 & & \\
$1^3D_1\to1^3P_2$ & 280.43 & 2.13 & 3.81 & 5.06 & & \\
$1^3D_2\to1^3P_1$ & 316.99 & 5.95 & 17.35 & 15.17 & & \\
$1^3D_2\to1^3P_2$ & 281.02 & 1.93 & 4.95 & 1.62 & & \\
$1^3D_3\to1^3P_2$ & 259.67 & 7.15 & 24.47 & 2.53 & & \\
$1^1D_2\to1^1P_1$ & 303.57 & 7.99 & 22.01 & & & \\
$2^3P_0\to2^3S_1$ & 166.35 & 5.93 & 11.75 & 13.68 & & \\
$2^3P_1\to2^3S_1$ & 257.99 & 3.20 & 16.94 & 1.73 & & \\
$2^3P_2\to2^3S_1$ & 301.25 & 15.22 & 17.76 & 22.86 & & \\
$2^1P_1\to2^1S_0$ & 343.37 & 2.75 & 21.25 & 2.52 & & \\
$2^3P_1\to1^3S_1$ & 276.10 & 26.61 & & & & \\
$2^3P_2\to1^3S_1$ & 296.18 & 26.29 & & & & \\
$2^1P_1\to1^1S_0$ & 53.98 & 0.019 & & & & \\
$2^3S_1\to1^3P_0$ & 308.83 & 1.042 & 0.743 & 1.29 & & \\
$2^3S_1\to1^3P_1$ & 219.22 & 0.377 & 0.894 & 0.02 & & \\
$2^3S_1\to1^3P_2$ & 182.68 & 1.88 & 1.22 & 1.89 & 1.59 & \\
$2^1S_0\to1^1P_1$ & 131.96 & 0.035 & 1.61 & 0.03 & 2.69 & \\
$3^3S_1\to2^3P_0$ & 288.89 & 0.019 & & & & \\
$3^3S_1\to2^3P_1$ & 197.99 & 0.0019 & & & & \\
$3^3S_1\to2^3P_2$ & 152.94 & 0.0027 & & & & \\
$3^3S_1\to1^3P_0$ & 70.08 & 13.39 & & & & \\
$3^3S_1\to1^3P_1$ & 62.48 & 23.53 & & & & \\
$3^3S_1\to1^3P_2$ & 59.28 & 29.40 & & & & \\
$3^1S_0\to2^1P_1$ & 125.19 & 0.131 & & & & \\
$3^1S_0\to1^1P_1$ & 68.70 & 11.143 & & & & \\
\hline\hline
\end{tabular*}
\end{table*}

\begin{table*}[htbp]
\caption{Magnetic dipole ($M1$) transition widths of the $D$ meson.}
\label{tab:M1D}
\begin{tabular*}{\textwidth}{@{\extracolsep{\fill}}lccccc@{}}
\hline\hline
Transition & $E_\gamma$ (MeV) & $\Gamma$ (keV) & \cite{Patel:2021aas} & \cite{Kher:2017wsq} & \cite{Devlani:2013kta} \\
\hline
$1^3S_1\to1^1S_0$ & 144.41 & 0.289 & 0.220 & 0.271 & 0.339 \\
$2^3S_1\to2^1S_0$ & 89.43 & 0.368 & 0.011 & 0.055 & 0.0007 \\
$2^3S_1\to1^1S_0$ & 664.19 & 23.431 & 5.782 & 6.371 & \\
$3^3S_1\to3^1S_0$ & 70.19 & 3.093 & 0.36 & 0.21 & \\
$1^1P_1\to1^3P_0$ & 100.85 & 3.2525 & 0.041 & 2.340 & \\
$1^3P_1\to1^1P_1$ & 85.14 & 2.97 & 0.001 & 0.314 & \\
\hline\hline
\end{tabular*}
\end{table*}

\begin{table*}[htbp]
\caption{Electric dipole ($E1$) transition widths of the $D_s$ meson.}
\label{tab:E1Ds}
\begin{tabular*}{\textwidth}{@{\extracolsep{\fill}}lccccccccc@{}}
\hline\hline
Transition & $E_\gamma$ (MeV) & $\Gamma$ (keV) & \cite{Chen:2020jku} & \cite{Radford:2009bs} & \cite{Green:2016occ} & \cite{Korner:1992pz} & \cite{Godfrey:2005ww} & \cite{Close:2005se} & \cite{Goity:2000dk} \\
\hline
$1^3P_0\to1^3S_1$ & 286.85 & 6.16 & 2.06 & 4.92 & 5.46 & & 1.9 & & $24.9\pm1.9$ \\
$1^3P_1\to1^3S_1$ & 356.55 & 7.7 & 4.79 & 15.5 & 17.4 & 5.6 & 4.41 & & 14.6 \\
$1^3P_2\to1^3S_1$ & 379.79 & 8.66 & 15.6 & 44.1 & 49.6 & 1.4 & 19.0 & 8.8 & 41.5 \\
$1^1P_1\to1^1S_0$ & 473.68 & 11.14 & 18.18 & 54.5 & 61.2 & 1.6 & 6.2 & 4.53 & 17.2 \\
$1^3D_1\to1^3P_0$ & 377.75 & 7.23 & 21.26 & & & & & & \\
$1^3D_1\to1^3P_1$ & 307.97 & 5.44 & 4.33 & & & & & & \\
$1^3D_1\to1^3P_2$ & 283.79 & 0.334 & 0.27 & & & & & & \\
$1^3D_2\to1^3P_1$ & 305.92 & 7.42 & 6.26 & & & & & & \\
$1^3D_2\to1^3P_2$ & 281.72 & 2.98 & 1.10 & & & & & & \\
$1^3D_3\to1^3P_2$ & 257.95 & 10.53 & 11.77 & & & & & & \\
$1^1D_2\to1^1P_1$ & 293.52 & 12.50 & 7.10 & & & & & & \\
$2^3P_0\to2^3S_1$ & 176.91 & 3.15 & 0.004 & & & & & & \\
$2^3P_1\to2^3S_1$ & 251.37 & 4.73 & 3.15 & & & & & & \\
$2^3P_2\to2^3S_1$ & 283.71 & 4.48 & 12.23 & & & & & & \\
$2^1P_1\to2^1S_0$ & 339.81 & 2.53 & 1.07 & & & & & & \\
$2^3P_0\to1^3S_1$ & 685.26 & 18.65 & 0.03 & & & & & & \\
$2^3P_1\to1^3S_1$ & 746.04 & 22.82 & 2.45 & & & & & & \\
$2^3P_2\to1^3S_1$ & 772.56 & 24.26 & 2.53 & & & & & & \\
$2^1P_1\to1^1S_0$ & 845.67 & 26.29 & 1.07 & & & & & & \\
$2^3S_1\to1^3P_0$ & 287.76 & 1.58 & 3.32 & 6.76 & 8.77 & & 2.4 & & \\
$2^3S_1\to1^3P_1$ & 215.36 & 3.31 & 0.30 & 0.24 & 0.41 & & 7.5 & & \\
$2^3S_1\to1^3P_2$ & 190.28 & 4.33 & 1.21 & 0.35 & 0.71 & & 7.6 & & \\
$2^1S_0\to1^1P_1$ & 125.64 & 2.86 & 0.07 & 0.01 & 0.05 & & 3.35 & & \\
$3^3S_1\to2^3P_0$ & 269.12 & 8.65 & 14.06 & & & & & & \\
$3^3S_1\to2^3P_1$ & 194.90 & 3.71 & 0.34 & & & & & & \\
$3^3S_1\to2^3P_2$ & 161.35 & 4.95 & 0.69 & & & & & & \\
$3^3S_1\to1^3P_0$ & 676.21 & 5.25 & 13.04 & & & & & & \\
$3^3S_1\to1^3P_1$ & 614.36 & 15.03 & 0.39 & & & & & & \\
$3^3S_1\to1^3P_2$ & 592.93 & 23.53 & 0.99 & & & & & & \\
$3^1S_0\to2^1P_1$ & 115.93 & 4.59 & 0.77 & & & & & & \\
$3^1S_0\to1^1P_1$ & 549.07 & 34.78 & 1.60 & & & & & & \\
\hline\hline
\end{tabular*}
\end{table*}

\begin{table*}[htbp]
\caption{Magnetic dipole ($M1$) transition widths of the $D_s$ meson.}
\label{tab:M1Ds}
\begin{tabular*}{\textwidth}{@{\extracolsep{\fill}}lcc@{}}
\hline\hline
Transition & $E_\gamma$ (MeV) & $\Gamma$ (keV) \\
\hline
$1^3S_1\to1^1S_0$ & 135.94 & 1.50 \\
$2^3S_1\to2^1S_0$ & 92.22 & 0.46 \\
$2^3S_1\to1^1S_0$ & 647.88 & 8.31 \\
$3^3S_1\to3^1S_0$ & 77.20 & 0.26 \\
$1^1P_1\to1^3P_0$ & 80.85 & 0.11 \\
$1^3P_1\to1^3P_0$ & 0.199 & 4.94 \\
$1^3D_2\to1^1D_2$ & 11.87 & 0.001 \\
$1^3D_2\to1^3D_3$ & 26.17 & 0.010 \\
$1^1D_2\to1^3D_3$ & 14.36 & 0.002 \\
\hline\hline
\end{tabular*}
\end{table*}

\subsection{Analysis of the electromagnetic transitions}
\label{sec:emanalysis}

The dominant radiative modes of the $1P$ for $D$ mesons given in Tab.~\ref{tab:E1D} are the $1^3P_J \to 1^3S_1\gamma$ and $1^1P_1 \to 1^1S_0\gamma$ transitions, which are obtained in $10.9 - 17.0$~keV range. These values are between the screened potential results of $8.7 - 25.5$~keV~\cite{Patel:2021aas} and the substantially larger widths of $17 - 51$~keV~\cite{Close:2005se}. The ordering $\Gamma(1^3P_0) < \Gamma(1^3P_1) < \Gamma(1^3P_2)$ being common to all models and driven purely by the growing photon phase space of the order of $\omega^3$.
The strong model scatter in individual entries e.g., the $1^3P_1\to1^3S_1$ width ranges over two orders of magnitude across the columns of Tab.~\ref{tab:E1D} which originates in the overlap integral of Eq.~\eqref{eq:ME1}, which involves cancellations between the oscillating Bessel functions and the nodal structure of the radial wave functions.
Thus, radiative widths are the single most discriminating wave function observable available, and a measurement of even one $1P\to1S\gamma$ branching fraction which is accessible at LHCb through $D_2^*(2460)\to D^*\gamma$ relative to $D^*\pi$ is expected to provide a decisive model test.

For the $D_s$ system (Tab.~\ref{tab:E1Ds}) the $E1$ widths are systematically smaller than their $D$ counterparts despite similar photon energies.
This is because the mean effective charge $\langle e_Q\rangle$ of Eq.~\eqref{eq:charges} suffers a partial cancellation between the charm and strange contributions which is absent for the charged-light-quark systems.
Radiative transitions are of prime importance precisely in this sector.
The observed decay chains of the $D_{s0}^*(2317)^\pm$ and $D_{s1}(2460)^\pm$ proceed through isospin violating $D_s^{(*)}\pi^0$ and radiative $D_s^{(*)}\gamma$ modes, so the ratios $\Gamma(D_s^*\gamma)/\Gamma(D_s\pi^0)$ measured by BaBar and Belle test any structural hypothesis for these states. 
Our $1^3P_0\to1^3S_1\gamma$ width of $6.2$~keV are comparable with the range $1.9-25$~keV covered by Refs.~\cite{Chen:2020jku,Radford:2009bs,Green:2016occ,Godfrey:2005ww,Goity:2000dk}.

Among the $M1$ transitions in Tab.~\ref{tab:M1D} and \ref{tab:M1Ds}, the ground state hyperfine transition $1^3S_1\to1^1S_0\gamma$ is phenomenologically important.
We obtain $0.289$~keV for $D^{*}\to D\gamma$ and other theoretical models predict values $0.22 - 0.34$~keV~\cite{Patel:2021aas,Kher:2017wsq,Devlani:2013kta}.
Combining with the measured total width $\Gamma(D^{*+}) = 83.4\pm1.8$~keV and radiative branching fraction $\mathcal{B}(D^{*+}\to D^+\gamma) = 1.6\%$~\cite{PDG2024}, the experimental radiative decay width is found to be $1.33\pm0.05$~keV for the $D^{*+}$.
Whereas, for the $D_s^{*}$ radiative mode dominates as the strong decays are isospin suppressed. Our predicted $D_s^*\to D_s\gamma$ width of $1.50$~keV is consistent with the total width upper limit $\Gamma(D_s^{*\pm}) < 1.9$~MeV \cite{PDG2024}.
The hindered transitions ($n\neq n'$), which disappear in the zero recoil limit and are generated entirely by wave function orthogonality breaking, show the largest model spread and should be regarded as order of magnitude estimates only. In that case, our $2^3S_1\to1^1S_0$ width is found to be 23.431~keV and others have predicted $5.782 - 6.371$~keV in Refs.~\cite{Patel:2021aas,Kher:2017wsq}.

\section{Strong decays}
\label{sec:strong}

The strong decays of the excited $D$ and $D_s$ states are analyzed within heavy quark effective theory (HQET). In the heavy quark limit $m_Q\to\infty$, the heavy quark spin decouples and the states are classified by the total angular momentum of the light degrees of freedom, $\bm{s}_\ell = \bm{s}_q + \bm{L}$, where $\bm{s}_q$ and $\bm{L}$ are the spin and orbital angular momentum of the light quark. Heavy-light mesons are then organized into doublets of total spin $J = s_\ell \pm \tfrac12$ and parity $P = (-1)^{L+1}$.

For $L=0$ ($S$ wave), $s_\ell^P = \frac12^-$ yields the doublet $(P,P^*)$ with $J^P = (0^-,1^-)$. For $L=1$ ($P$ wave) there are two doublets: $s_\ell^P = \frac12^+$ with $(P_0^*,P_1')$, $J^P = (0^+,1^+)$, and $s_\ell^P = \frac32^+$ with $(P_1,P_2^*)$, $J^P = (1^+,2^+)$.
For $L=2$ ($D$ wave), $s_\ell^P = \frac32^-$ with $(P_1^*,P_2)$, $J^P = (1^-,2^-)$, and $s_\ell^P = \frac52^-$ with $(P_2',P_3^*)$, $J^P = (2^-,3^-)$.
For $L=3$ ($F$ wave), $s_\ell^P = \tfrac52^+$ with $(P_2'^*,P_3)$, $J^P = (2^+,3^+)$, and $s_\ell^P = \tfrac72^+$ with $(P_3',P_4^*)$, $J^P = (3^+,4^+)$.
These spin doublets are described by the superfields $H_a$, $S_a$, $T_a$, $X_a$, $Y_a$, $Z_a$, and $R_a$ of heavy meson effective theory~\cite{Wang:2013tka},
\begin{eqnarray}
H_a &=&\ \frac{1+\slashed{v}}{2}\left[P^{*\mu}_{a}\gamma_\mu - P_a\gamma_5\right],\nonumber\\
S_a &=& \frac{1+\slashed{v}}{2}\left[P^{\mu}_{1a}\gamma_\mu\gamma_5 - P^*_{0a}\right], \nonumber\\
T^{\mu}_a &=&\ \frac{1+\slashed{v}}{2}\left\{P^{*\mu\nu}_{2a}\gamma_\nu - P_{1a\nu}\sqrt{\tfrac{3}{2}} \gamma_5\left[g^{\mu\nu}-\frac{\gamma^\nu(\gamma^\mu-v^\mu)}{3}\right]\right\}, \nonumber\\
X^{\mu}_a &=&\ \frac{1+\slashed{v}}{2}\left\{P^{\mu\nu}_{2a}\gamma_5\gamma_\nu - P^{*}_{1a\nu}\sqrt{\tfrac{3}{2}}\left[g^{\mu\nu}-\frac{\gamma^\nu(\gamma^\mu+v^\mu)}{3}\right]\right\},\nonumber\\
Y^{\mu\nu}_a &=&\ \frac{1+\slashed{v}}{2}\left\{P^{*\mu\nu\sigma}_{3a}\gamma_\sigma - P^{\alpha\beta}_{2a}\sqrt{\tfrac{5}{3}}\,\gamma_5\left[g^{\mu}_{\alpha}g^{\nu}_{\beta}-\frac{g^{\nu}_{\beta}\gamma_\alpha(\gamma^\mu-v^\mu)}{5}-\frac{g^{\mu}_{\alpha}\gamma_\beta(\gamma^\nu-v^\nu)}{5}\right]\right\}, \nonumber\\
Z^{\mu\nu}_a &=&\ \frac{1+\slashed{v}}{2}\left\{P^{\mu\nu\sigma}_{3a}\gamma_5\gamma_\sigma - P^{*\alpha\beta}_{2a}\sqrt{\tfrac{5}{3}}\left[g^{\mu}_{\alpha}g^{\nu}_{\beta}-\frac{g^{\nu}_{\beta}\gamma_\alpha(\gamma^\mu+v^\mu)}{5}-\frac{g^{\mu}_{\alpha}\gamma_\beta(\gamma^\nu+v^\nu)}{5}\right]\right\}, \nonumber\\
R^{\mu\nu\rho}_a &=&\ \frac{1+\slashed{v}}{2}\Bigg\{P^{*\mu\nu\rho\sigma}_{4a}\gamma_5\gamma_\sigma - P^{\alpha\beta\tau}_{3a}\sqrt{\tfrac{7}{4}}\bigg[g^{\mu}_{\alpha}g^{\nu}_{\beta}g^{\rho}_{\tau}-\frac{g^{\nu}_{\beta}g^{\rho}_{\tau}\gamma_\alpha(\gamma^\mu-v^\mu)}{7} -\frac{g^{\mu}_{\alpha}g^{\rho}_{\tau}\gamma_\beta(\gamma^\nu-v^\nu)}{7}-\frac{g^{\mu}_{\alpha}g^{\nu}_{\beta}\gamma_\tau(\gamma^\rho-v^\rho)}{7}\bigg]\Bigg\},\nonumber \\
\label{eq:superfields}
\end{eqnarray}
where $a$ labels the light quark flavor ($u/d$ or $s$), $v$ is the meson four velocity which is conserved in strong interactions, and the field operators $P^{(*)}$ contain a factor $\sqrt{m_Q}$ having mass dimension $3/2$. The octet of light pseudoscalar mesons enters through $\xi = e^{i\mathcal{M}/f_\pi}$ with
\begin{equation}
\mathcal{M} =
\begin{pmatrix}
\frac{\pi^0}{\sqrt{2}}+\frac{\eta}{\sqrt{6}} & \pi^+ & K^+ \\[4pt]
\pi^- & -\frac{\pi^0}{\sqrt{2}}+\frac{\eta}{\sqrt{6}} & K^0 \\[4pt]
K^- & \bar{K}^0 & -\sqrt{\tfrac{2}{3}}\,\eta
\end{pmatrix}.
\label{eq:Mmatrix}
\end{equation}

The effective heavy meson chiral Lagrangians describing two body strong decays by emission of a light pseudoscalar meson are~\cite{Wang:2013tka}
\begin{eqnarray}
\mathcal{L}_H &=& g_H\,{\rm Tr}\!\left[\bar{H}_a H_b \gamma_\mu\gamma_5 A^{\mu}_{ba}\right], \nonumber\\
\mathcal{L}_S &=& g_S\,{\rm Tr}\!\left[\bar{H}_a S_b \gamma_\mu\gamma_5 A^{\mu}_{ba}\right] + {\rm H.c.}, \nonumber\\
\mathcal{L}_T &=& \frac{g_T}{\Lambda_\chi}\,{\rm Tr}\!\left[\bar{H}_a T^{\mu}_b \left(i D_\mu \slashed{A} + i\slashed{D} A_\mu\right)_{ba}\gamma_5\right] + {\rm H.c.}, \nonumber\\
\mathcal{L}_X &=& \frac{g_X}{\Lambda_\chi}\,{\rm Tr}\!\left[\bar{H}_a X^{\mu}_b \left(i D_\mu \slashed{A} + i\slashed{D} A_\mu\right)_{ba}\gamma_5\right] + {\rm H.c.}, \nonumber\\
\mathcal{L}_Y &=& \frac{1}{\Lambda_\chi^2}\,{\rm Tr}\!\left[\bar{H}_a Y^{\mu\nu}_b \left(k_1^Y \{D_\mu,D_\nu\} A_\lambda + k_2^Y (D_\mu D_\lambda A_\nu + D_\nu D_\lambda A_\mu)\right)_{ba}\gamma^\lambda\gamma_5\right] + {\rm H.c.}, \nonumber\\
\mathcal{L}_Z &=& \frac{1}{\Lambda_\chi^2}\,{\rm Tr}\!\left[\bar{H}_a Z^{\mu\nu}_b \left(k_1^Z \{D_\mu,D_\nu\} A_\lambda + k_2^Z (D_\mu D_\lambda A_\nu + D_\nu D_\lambda A_\mu)\right)_{ba}\gamma^\lambda\gamma_5\right] + {\rm H.c.}, \nonumber\\
\mathcal{L}_R &=& \frac{1}{\Lambda_\chi^3}\,{\rm Tr}\!\left[\bar{H}_a R^{\mu\nu\rho}_b \left(k_1^R \{D_\mu,D_\nu,D_\rho\} A_\lambda + k_2^R\left(\{D_\mu,D_\rho\} D_\lambda A_\nu + \{D_\nu,D_\rho\} D_\lambda A_\mu + \{D_\mu,D_\nu\} D_\lambda A_\rho\right)\right)_{ba}\gamma^\lambda\gamma_5\right] + {\rm H.c.},\nonumber \\
\label{eq:lagrangians}
\end{eqnarray}
where the vector and axial vector combinations of the pseudo Goldstone fields are
\begin{eqnarray}
\mathcal{V}_{\mu ba} &=& \frac{1}{2}\left(\xi^\dagger\partial_\mu\xi + \xi\,\partial_\mu\xi^\dagger\right)_{ba},\nonumber \\
A_{\mu ba} &=& \frac{i}{2}\left(\xi^\dagger\partial_\mu\xi - \xi\,\partial_\mu\xi^\dagger\right)_{ba},
\label{eq:currents}
\end{eqnarray}
the chirally covariant derivative is $D_{\mu ba} = -\delta_{ba}\partial_\mu + \mathcal{V}_{\mu ba}$, the anticommutators are defined as $\{D_\mu,D_\nu\} = D_\mu D_\nu + D_\nu D_\mu$ and $\{D_\mu,D_\nu,D_\rho\} = \sum_{\rm perms} D_\mu D_\nu D_\rho$, and $\Lambda_\chi = 1$~GeV is the chiral symmetry breaking scale. The coupling constants $g_H$, $g_S$, $g_T$, $g_X$, $g_Y$, $g_Z$, and $g_R$ control the decays of the corresponding doublets.

The partial width for a two body strong decay to a final heavy meson plus a light pseudoscalar meson $\mathcal{P}$ is
\begin{equation}
\Gamma = \frac{1}{2J+1}\sum \frac{p_f}{8\pi M_i^2}\,|T|^2 ,
\label{eq:gammastrong}
\end{equation}
with the final state momentum
\begin{equation}
p_f = \frac{\sqrt{\left[M_i^2-(M_f+m_{\mathcal{P}})^2\right]\left[M_i^2-(M_f-m_{\mathcal{P}})^2\right]}}{2M_i},
\label{eq:pf}
\end{equation}
where $T$ is the scattering amplitude, $i$ and $f$ label the initial and final heavy mesons, $J$ is the total angular momentum of the initial meson, and the sum runs over the polarizations of the final states. The light pseudoscalar masses are adapted from the PDG~\cite{PDG2024} as $M_{\pi^\pm} = 139.57$~MeV, $M_{\pi^0} = 134.98$~MeV, $M_{K^\pm} = 493.68$~MeV, $M_{K^0} = 497.61$~MeV, and $M_{\eta} = 547.86$~MeV, while the $D$ and $D_s$ masses are from Tab. \ref{tab:massD_SP} - \ref{tab:massDs_DF}. The flavour coefficients are $C_{\pi^\pm} = C_{K^\pm} = C_{K^0} = C_{\bar{K}^0} = 1$, $C_{\pi^0} = \frac12$, and $C_{\eta} = \frac23$ (or $\frac16$), as appropriate to the light quark content.

\subsection{Isospin violation in the $D_s$ sector}
\label{sec:isospin}

The charm-strange mesons are isoscalar ($I=0$) states, so every transition of the type $D_s^{*} \to D_s^{(*)}\pi^0$ changes isospin by one unit and is strictly forbidden in the isospin symmetric limit. Such modes are called isospin-breaking modes and proceed only through $\pi^0-\eta$ mixing. These modes are controlled by the difference of the current light quark masses. Consequently the corresponding partial widths obtained from Eq.~\eqref{eq:gammastrong} must be multiplied by the square of the suppression factor \cite{Colangelo:2003vg, Matsuki:2011xp,Gandhi:2022nnk},
\begin{equation}
\epsilon^{2} = \frac{3}{16}\left(\frac{m_d - m_u}{m_s - \frac{m_u+m_d}{2}}\right)^{2} \simeq 1.5\times10^{-4},
\label{eq:isospin}
\end{equation}
where $m_u$, $m_d$ and $m_s$ are the current quark masses. This is the charm-sector analogue of the mechanism that governs the isospin violating decay of the strange-bottom scalar meson, and it is the reason why the $D_{s0}^*(2317)^\pm$ and $D_{s1}(2460)^\pm$, which lie below the $DK$ and $D^*K$ thresholds, are extremely narrow states even though their open hadronic channels are $S$-wave transitions with sizeable phase space.

Throughout this work every $D_s^{(*)}\pi^0$ entry of the present calculation, i.e. in Tab.~\ref{tab:strongDs1}, \ref{tab:strongDs2}, \ref{tab:ratiosDs1}, and \ref{tab:ratiosDs2}, already includes the factor $\epsilon^2$ of Eq.~\eqref{eq:isospin}. The values quoted from the literature in the comparison columns are reproduced as published and are not rescaled. Isospin conserving channels, i.e. $DK$, $D^*K$, $D_s^{(*)}\eta$, and all channels of the non-strange $D$ mesons, are unaffected.

The explicit partial width expressions for each doublet transition $(J^P,J'^P)_{s_\ell} \to (0^-,1^-)_{1/2} + \mathcal{P}$ have been derived in Ref.~\cite{Wang:2013tka}. These decay widths reduce to Eq.~\eqref{eq:gammastrong} with amplitudes scaling as $|T|^2 \propto g^2\, p_f^{2\ell}\,(p_f^2 + m_{\mathcal{P}}^2)^{k}/\Lambda_\chi^{2(\ell+k-1)}$, where $\ell$ is the orbital angular momentum of the emitted pseudoscalar ($\ell = 1$ for $S$-wave initial doublets up to $\ell = 4$ for the $s_\ell = \frac72$ $F$-wave doublet) and $k = 0$ or $1$ according to the chiral structure of the vertex.
We employ these expressions of Ref.~\cite{Wang:2013tka} throughout the study. The strong $p_f^{2\ell+1}$ threshold behavior they encode makes the width patterns much sensitive to the doublet assignment.

Since the coupling constants of the effective Lagrangians are not fixed within the model, all partial widths are quoted in units of the appropriate coupling squared ($g_H^2$, $g_S^2$, $g_T^2$, $g_X^2$, $g_Y^2$, $g_Z^2$, $g_R^2$). Ratios of partial widths within a given doublet transition are therefore parameter free predictions.

Using this framework, we compute the strong decay widths of the $2S$ and $1P$ states $D_0(2550)^0$, $D_1^*(2600)^0$, $D_0^*(2300)^0$, $D_1(2420)^0$, and $D_2^*(2460)$ in Tab.~\ref{tab:strongD1} and \ref{tab:strongD2}, of the $1D$ states $D_1^*(2760)^0$, $D_2(2740)^0$, and $D_3^*(2750)$ in Tab.~\ref{tab:strongD2} and \ref{tab:strongD3}, and of the $D_J(3000)^0$ and $D_J^*(3000)^0$ states under the various competing assignments in Tab.~\ref{tab:strongD4} and \ref{tab:strongD5}. The $D_0(2550)^0$ is treated as the $2^1S_0$ state with $J^P = 0^-$, the $D_0^*(2300)^0$ as the $1^3P_0$ state with $J^P = 0^+$, and the $D_1(2420)^0$ receives contributions from both the $s_\ell = \frac12$ and $s_\ell = \frac32$ $1P$ doublets ($1^1P_1-1^3P_1$ mixing), for which both possibilities are tabulated. Likewise, for the $D_2(2740)^0$, both the $s_\ell^P = \tfrac32^-$ and $\tfrac52^-$ $1D$ assignments are examined. The width patterns of the two hypotheses differ noticeably (Tab.~\ref{tab:strongD3}), so precise measurements of the ratios in Tab.~\ref{tab:ratiosD2} can discriminate between them.

\begin{table*}[htbp]
\caption{Strong decay widths of the $2S$ and $1P$ $D$-meson states, in units of the squared couplings. \label{tab:strongD1}}
\begin{tabular*}{\textwidth}{@{\extracolsep{\fill}}llllllllll@{}}
\hline\hline
Meson & $n$ & $s_\ell$ & $J^P$ & State & Channel & Present & \cite{Wang:2013tka} & \cite{LHCb:2016lxy} & \cite{LHCb:2013jjb} \\
\hline
$D_0(2550)^0$ & 2 & $\frac12$ & $0^-$ & $2^1S_0$ & $D^{*+}\pi^-$ & $0.862g_H^2$ & $0.867g_H^2$ & $0.865g_H^2$ & $0.709g_H^2$ \\
 & & & & & $D^{*0}\pi^0$ & $0.443g_H^2$ & $0.442g_H^2$ & $0.442g_H^2$ & $0.363g_H^2$ \\
 & & & & & $D^{*0}\eta$ & $\ldots$ & $0.006g_H^2$ & $0.015g_H^2$ & $0.004g_H^2$ \\
$D_1^*(2600)^0$ & 2 & $\frac12$ & $1^-$ & $2^3S_1$ & $D^{+}\pi^-$ & $0.674g_H^2$ & $0.619g_H^2$ & $0.680g_H^2$ & $0.541g_H^2$ \\
 & & & & & $D^{0}\pi^0$ & $0.369g_H^2$ & $0.314g_H^2$ & $0.346g_H^2$ & $0.275g_H^2$ \\
 & & & & & $D_s^{+}K^-$ & $\ldots$ & $0.168g_H^2$ & $0.156g_H^2$ & $0.199g_H^2$ \\
 & & & & & $D^{0}\eta$ & $\ldots$ & $0.151g_H^2$ & $0.157g_H^2$ & $0.048g_H^2$ \\
 & & & & & $D^{*+}\pi^-$ & $0.751g_H^2$ & $0.784g_H^2$ & $0.887g_H^2$ & $0.782g_H^2$ \\
 & & & & & $D^{*0}\pi^0$ & $0.377g_H^2$ & $0.399g_H^2$ & $0.451g_H^2$ & $0.398g_H^2$ \\
 & & & & & $D_s^{*+}K^-$ & $\ldots$ & $0.079g_H^2$ & $0.033g_H^2$ & $0.078g_H^2$ \\
 & & & & & $D^{*0}\eta$ & $\ldots$ & $0.062g_H^2$ & $0.079g_H^2$ & $0.031g_H^2$ \\
$D_0^*(2300)^0$ & 1 & $\frac12$ & $0^+$ & $1^3P_0$ & $D^{+}\pi^-$ & $0.486g_S^2$ & & & \\
 & & & & & $D^{0}\pi^0$ & $0.243g_S^2$ & & & \\
$D_1(2420)^0$ & 1 & $\frac12$ & $1^+$ & $1P_1'$ & $D^{*+}\pi^-$ & $2.892g_S^2$ & & & \\
 & & & & & $D^{*0}\pi^0$ & $1.446g_S^2$ & & & \\
 & & & & & $D_s^{*+}K^-$ & $2.034g_S^2$ & & & \\
 & & & & & $D^{*0}\eta$ & $0.436g_S^2$ & & & \\
$D_1(2420)^0$ & 1 & $\frac32$ & $1^+$ & $1P_1$ & $D^{*+}\pi^-$ & $0.054g_T^2$ & $0.056g_T^2$ & $0.057g_T^2$ & \\
 & & & & & $D^{*0}\pi^0$ & $0.027g_T^2$ & $0.029g_T^2$ & $0.030g_T^2$ & \\
$D_2^*(2460)^0$ & 1 & $\frac32$ & $2^+$ & $1^3P_2$ & $D^{+}\pi^-$ & $0.123g_T^2$ & $0.128g_T^2$ & $0.125g_T^2$ & $0.126g_T^2$ \\
 & & & & & $D^{0}\pi^0$ & $0.062g_T^2$ & $0.067g_T^2$ & $0.065g_T^2$ & $0.066g_T^2$ \\
 & & & & & $D_s^{+}K^-$ & $\ldots$ & $0.006g_T^2$ & & \\
 & & & & & $D^{0}\eta$ & $\ldots$ & $0.0006g_T^2$ & & \\
 & & & & & $D^{*+}\pi^-$ & $0.0524g_T^2$ & $0.056g_T^2$ & $0.054g_T^2$ & $0.056g_T^2$ \\
 & & & & & $D^{*0}\pi^0$ & $0.032g_T^2$ & $0.030g_T^2$ & $0.030g_T^2$ & $0.029g_T^2$ \\
 \hline\hline
\end{tabular*}
\end{table*}

\begin{table*}[htbp]
\caption{Strong decay widths of the charged $D_2^*(2460)$ and of the $1D$ $D$-meson states (continued from Tab.~\ref{tab:strongD1}), in units of the squared couplings. \label{tab:strongD2}}
\begin{tabular*}{\textwidth}{@{\extracolsep{\fill}}llllllll@{}}
\hline\hline
Meson & $n$ & $s_\ell$ & $J^P$ & State & Channel & Present & \cite{Wang:2013tka} \\
\hline
$D_2^*(2460)^+$ & 1 & $\frac32$ & $2^+$ & $1^3P_2$ & $D^{0}\pi^+$ & $0.110g_T^2$ & $0.132g_T^2$ \\
 & & & & & $D^{+}\pi^0$ & $0.058g_T^2$ & $0.064g_T^2$ \\
 & & & & & $D^{*0}\pi^+$ & $0.052g_T^2$ & $0.058g_T^2$ \\
 & & & & & $D^{*+}\pi^0$ & $0.022g_T^2$ & $0.028g_T^2$ \\
$D_2^*(2460)^-$ & 1 & $\frac32$ & $2^+$ & $1^3P_2$ & $D^{0}\pi^-$ & $0.137g_T^2$ & \\
 & & & & & $D^{-}\pi^0$ & $0.066g_T^2$ & \\
 & & & & & $D^{*0}\pi^-$ & $0.056g_T^2$ & $0.062g_T^2$ \\
 & & & & & $D^{*-}\pi^0$ & $0.026g_T^2$ & $0.030g_T^2$ \\
$D_1^*(2760)^0$ & 1 & $\frac32$ & $1^-$ & $1^3D_1$ & $D^{+}\pi^-$ & $1.276g_X^2$ & $1.285g_X^2$ \\
 & & & & & $D^{0}\pi^0$ & $0.652g_X^2$ & $0.656g_X^2$ \\
 & & & & & $D_s^{+}K^-$ & $0.448g_X^2$ & $0.459g_X^2$ \\
 & & & & & $D^{0}\eta$ & $0.493g_X^2$ & $0.501g_X^2$ \\
 & & & & & $D^{*+}\pi^-$ & $0.345g_X^2$ & $0.335g_X^2$ \\
 & & & & & $D^{*0}\pi^0$ & $0.171g_X^2$ & $0.171g_X^2$ \\
 & & & & & $D_s^{*+}K^-$ & $0.062g_X^2$ & $0.061g_X^2$ \\
 & & & & & $D^{*0}\eta$ & $0.085g_X^2$ & $0.088g_X^2$ \\
\hline\hline
\end{tabular*}
\end{table*}

\begin{table*}[htbp]
\caption{Strong decay widths of the $D_2(2740)^0$ under the two possible $1D$ doublet assignments, and of the $D_3^*(2750)$, in units of the squared couplings. \label{tab:strongD3}}
\begin{tabular*}{\textwidth}{@{\extracolsep{\fill}}llllllllll@{}}
\hline\hline
Meson & $n$ & $s_\ell$ & $J^P$ & State & Channel & Present & \cite{Wang:2013tka} & \cite{LHCb:2016lxy} & \cite{LHCb:2013jjb} \\
\hline
$D_2(2740)^0$ & 1 & $\frac32$ & $2^-$ & $1D_2$ & $D^{*+}\pi^-$ & $0.129g_X^2$ & $0.875g_X^2$ & & \\
 & & & & & $D^{*0}\pi^0$ & $0.042g_X^2$ & $0.447g_X^2$ & & \\
 & & & & & $D_s^{*+}K^-$ & $0.001g_X^2$ & $0.133g_X^2$ & & \\
 & & & & & $D^{*0}\eta$ & $0.002g_X^2$ & $0.207g_X^2$ & & \\
$D_2(2740)^0$ & 1 & $\frac52$ & $2^-$ & $1D_2'$ & $D^{*+}\pi^-$ & $0.138g_Y^2$ & $0.127g_Y^2$ & $0.127g_Y^2$ & \\
 & & & & & $D^{*0}\pi^0$ & $0.070g_Y^2$ & $0.066g_Y^2$ & $0.066g_Y^2$ & \\
 & & & & & $D_s^{*+}K^-$ & $0.002g_Y^2$ & $0.001g_Y^2$ & $0.002g_Y^2$ & \\
 & & & & & $D^{*0}\eta$ & $0.004g_Y^2$ & $0.002g_Y^2$ & $0.001g_Y^2$ & \\
$D_3^*(2750)^0$ & 1 & $\frac52$ & $3^-$ & $1^3D_3$ & $D^{+}\pi^-$ & $0.179g_Y^2$ & $0.173g_Y^2$ & $0.191g_Y^2$ & $0.176g_Y^2$ \\
 & & & & & $D^{0}\pi^0$ & $0.085g_Y^2$ & $0.089g_Y^2$ & $0.099g_Y^2$ & $0.091g_Y^2$ \\
 & & & & & $D_s^{+}K^-$ & $0.019g_Y^2$ & $0.017g_Y^2$ & $0.021g_Y^2$ & $0.018g_Y^2$ \\
 & & & & & $D^{0}\eta$ & $0.005g_Y^2$ & $0.024g_Y^2$ & $0.007g_Y^2$ & $0.0066g_Y^2$ \\
 & & & & & $D^{*+}\pi^-$ & $0.088g_Y^2$ & $0.089g_Y^2$ & $0.100g_Y^2$ & $0.090g_Y^2$ \\
 & & & & & $D^{*0}\pi^0$ & $0.045g_Y^2$ & $0.046g_Y^2$ & $0.052g_Y^2$ & $0.047g_Y^2$ \\
 & & & & & $D_s^{*+}K^-$ & $0.005g_Y^2$ & $0.002g_Y^2$ & $0.003g_Y^2$ & $0.002g_Y^2$ \\
 & & & & & $D^{*0}\eta$ & $0.004g_Y^2$ & $0.004g_Y^2$ & $0.002g_Y^2$ & $0.001g_Y^2$ \\
$D_3^*(2750)^+$ & 1 & $\frac52$ & $3^-$ & $1^3D_3$ & $D^{0}\pi^+$ & $0.186g_Y^2$ & $0.191g_Y^2$ & $0.189g_Y^2$ & \\
 & & & & & $D^{+}\pi^0$ & $0.088g_Y^2$ & $0.093g_Y^2$ & $0.092g_Y^2$ & \\
 & & & & & $D_s^{+}K^0$ & $0.019g_Y^2$ & $0.019g_Y^2$ & $0.018g_Y^2$ & \\
 & & & & & $D^{+}\eta$ & $0.005g_Y^2$ & $0.006g_Y^2$ & $0.006g_Y^2$ & \\
 & & & & & $D^{*0}\pi^+$ & $0.092g_Y^2$ & $0.099g_Y^2$ & $0.098g_Y^2$ & \\
 & & & & & $D^{*+}\pi^0$ & $0.044g_Y^2$ & $0.049g_Y^2$ & $0.048g_Y^2$ & \\
 & & & & & $D_s^{*+}K^0$ & $0.005g_Y^2$ & $0.002g_Y^2$ & $0.002g_Y^2$ & \\
 & & & & & $D^{*+}\eta$ & $0.002g_Y^2$ & $0.001g_Y^2$ & $0.001g_Y^2$ & \\
$D_3^*(2750)^-$ & 1 & $\frac52$ & $3^-$ & $1^3D_3$ & $D^{0}\pi^-$ & $0.216g_Y^2$ & $0.226g_Y^2$ & & \\
 & & & & & $D^{-}\pi^0$ & $0.102g_Y^2$ & $0.111g_Y^2$ & & \\
 & & & & & $D_s^{-}K^0$ & $0.019g_Y^2$ & $0.027g_Y^2$ & & \\
 & & & & & $D^{-}\eta$ & $0.005g_Y^2$ & $0.008g_Y^2$ & & \\
 & & & & & $D^{*-}\pi^-$ & $0.108g_Y^2$ & $0.122g_Y^2$ & & \\
 & & & & & $D^{*-}\pi^0$ & $0.054g_Y^2$ & $0.060g_Y^2$ & & \\
 & & & & & $D_s^{*-}K^0$ & $0.005g_Y^2$ & $0.004g_Y^2$ & & \\
 & & & & & $D^{*-}\eta$ & $0.002g_Y^2$ & $0.002g_Y^2$ & & \\
 \hline\hline
\end{tabular*}
\end{table*}

\begin{table*}[htbp]
\caption{Strong decay widths of the $D_J(3000)^0$ and $D_J^*(3000)^0$ states under the competing $1F$ and $3S$ assignments, in units of the squared couplings. \label{tab:strongD4}}
\begin{tabular*}{\textwidth}{@{\extracolsep{\fill}}lllllllll@{}}
\hline\hline
Meson & $n$ & $s_\ell$ & $J^P$ & State & Channel & Present & \cite{Wang:2013tka} & \cite{LHCb:2013jjb} \\
\hline
$D_2^*(3000)^0$ & 1 & $\frac52$ & $2^+$ & $1^3F_2$ & $D^{+}\pi^-$ & $1.029g_Z^2$ & $1.034g_Z^2$ & $1.031g_Z^2$ \\
 & & & & & $D^{0}\pi^0$ & $0.532g_Z^2$ & $0.529g_Z^2$ & $0.527g_Z^2$ \\
 & & & & & $D_s^{+}K^-$ & $0.384g_Z^2$ & $0.387g_Z^2$ & $0.381g_Z^2$ \\
 & & & & & $D^{0}\eta$ & $0.106g_Z^2$ & $0.407g_Z^2$ & $0.102g_Z^2$ \\
 & & & & & $D^{*+}\pi^-$ & $0.354g_Z^2$ & $0.355g_Z^2$ & $0.355g_Z^2$ \\
 & & & & & $D^{*0}\pi^0$ & $0.181g_Z^2$ & $0.182g_Z^2$ & $0.181g_Z^2$ \\
 & & & & & $D_s^{*+}K^-$ & $0.089g_Z^2$ & $0.092g_Z^2$ & $0.092g_Z^2$ \\
 & & & & & $D^{*0}\eta$ & $0.099g_Z^2$ & $0.112g_Z^2$ & $0.028g_Z^2$ \\
$D_J(3000)^0$ & 1 & $\frac52$ & $3^+$ & $1F_3$ & $D^{*+}\pi^-$ & $0.718g_Z^2$ & $0.716g_Z^2$ & \\
 & & & & & $D^{*0}\pi^0$ & $0.369g_Z^2$ & $0.365g_Z^2$ & \\
 & & & & & $D_s^{*+}K^-$ & $0.168g_Z^2$ & $0.165g_Z^2$ & \\
 & & & & & $D^{*0}\eta$ & $0.199g_Z^2$ & $0.210g_Z^2$ & \\
$D_J(3000)^0$ & 1 & $\frac72$ & $3^+$ & $1F_3'$ & $D^{*+}\pi^-$ & $1.693g_R^2$ & $1.703g_R^2$ & \\
 & & & & & $D^{*0}\pi^0$ & $0.869g_R^2$ & $0.877g_R^2$ & \\
 & & & & & $D_s^{*+}K^-$ & $0.135g_R^2$ & $0.136g_R^2$ & \\
 & & & & & $D^{*0}\eta$ & $0.214g_R^2$ & $0.210g_R^2$ & \\
$D_4^*(3000)^0$ & 1 & $\frac72$ & $4^+$ & $1^3F_4$ & $D^{+}\pi^-$ & $2.416g_R^2$ & $2.422g_R^2$ & $2.415g_R^2$ \\
 & & & & & $D^{0}\pi^0$ & $1.249g_R^2$ & $1.250g_R^2$ & $1.246g_R^2$ \\
 & & & & & $D_s^{+}K^-$ & $0.432g_R^2$ & $0.439g_R^2$ & $0.426g_R^2$ \\
 & & & & & $D^{0}\eta$ & $0.184g_R^2$ & $0.520g_R^2$ & $0.129g_R^2$ \\
 & & & & & $D^{*+}\pi^-$ & $1.258g_R^2$ & $1.257g_R^2$ & $1.254g_R^2$ \\
 & & & & & $D^{*0}\pi^0$ & $0.644g_R^2$ & $0.647g_R^2$ & $0.646g_R^2$ \\
 & & & & & $D_s^{*+}K^-$ & $0.121g_R^2$ & $0.123g_R^2$ & $0.123g_R^2$ \\
 & & & & & $D^{*0}\eta$ & $0.084g_R^2$ & $0.179g_R^2$ & $0.045g_R^2$ \\
$D_J(3000)^0$ & 3 & $\frac12$ & $0^-$ & $3^1S_0$ & $D^{*+}\pi^-$ & $3.807g_H^2$ & $3.226g_H^2$ & $3.217g_H^2$ \\
 & & & & & $D^{*0}\pi^0$ & $1.907g_H^2$ & $1.627g_H^2$ & $1.623g_H^2$ \\
 & & & & & $D_s^{*+}K^-$ & $1.669g_H^2$ & $1.438g_H^2$ & $1.435g_H^2$ \\
 & & & & & $D^{*0}\eta$ & $1.563g_H^2$ & $1.225g_H^2$ & $0.306g_H^2$ \\
 \hline\hline
\end{tabular*}
\end{table*}

\begin{table*}[htbp]
\caption{Strong decay widths of the $D_J(3000)^0$ and $D_J^*(3000)^0$ states under the $3S$ and $2P$ assignments (continued from Tab.~\ref{tab:strongD4}), in units of the squared couplings. \label{tab:strongD5}}
\begin{tabular*}{\textwidth}{@{\extracolsep{\fill}}lllllllll@{}}
\hline\hline
Meson & $n$ & $s_\ell$ & $J^P$ & State & Channel & Present & \cite{Wang:2013tka} & \cite{LHCb:2013jjb} \\
\hline
$D_J^*(3000)^0$ & 3 & $\frac12$ & $1^-$ & $3^3S_1$ & $D^{+}\pi^-$ & $1.844g_H^2$ & $1.498g_H^2$ & \\
 & & & & & $D^{0}\pi^0$ & $0.923g_H^2$ & $0.755g_H^2$ & \\
 & & & & & $D_s^{+}K^-$ & $1.183g_H^2$ & $0.877g_H^2$ & \\
 & & & & & $D^{0}\eta$ & $0.896g_H^2$ & $0.683g_H^2$ & \\
 & & & & & $D^{*+}\pi^-$ & $2.965g_H^2$ & $2.346g_H^2$ & \\
 & & & & & $D^{*0}\pi^0$ & $1.480g_H^2$ & $1.182g_H^2$ & \\
 & & & & & $D_s^{*+}K^-$ & $2.111g_H^2$ & $1.119g_H^2$ & \\
 & & & & & $D^{*0}\eta$ & $1.300g_H^2$ & $0.934g_H^2$ & \\
$D_J^*(3000)^0$ & 2 & $\frac12$ & $0^+$ & $2^3P_0$ & $D^{+}\pi^-$ & $4.055g_S^2$ & $4.598g_S^2$ & \\
 & & & & & $D^{0}\pi^0$ & $2.327g_S^2$ & $2.315g_S^2$ & \\
 & & & & & $D_s^{+}K^-$ & $3.662g_S^2$ & $3.763g_S^2$ & \\
 & & & & & $D^{0}\eta$ & $2.833g_S^2$ & $2.993g_S^2$ & \\
$D_J(3000)^0$ & 2 & $\frac12$ & $1^+$ & $2P_1'$ & $D^{*+}\pi^-$ & $3.312g_S^2$ & $3.325g_S^2$ & $3.315g_S^2$ \\
 & & & & & $D^{*0}\pi^0$ & $1.646g_S^2$ & $1.674g_S^2$ & $1.670g_S^2$ \\
 & & & & & $D_s^{*+}K^-$ & $2.407g_S^2$ & $2.145g_S^2$ & $2.409g_S^2$ \\
 & & & & & $D^{*0}\eta$ & $0.436g_S^2$ & $2.067g_S^2$ & $0.515g_S^2$ \\
$D_J(3000)^0$ & 2 & $\frac32$ & $1^+$ & $2P_1$ & $D^{*+}\pi^-$ & $2.722g_T^2$ & $2.733g_T^2$ & $2.725g_T^2$ \\
 & & & & & $D^{*0}\pi^0$ & $1.473g_T^2$ & $1.389g_T^2$ & $1.385g_T^2$ \\
 & & & & & $D_s^{*+}K^-$ & $0.695g_T^2$ & $0.688g_T^2$ & $0.725g_T^2$ \\
 & & & & & $D^{*0}\eta$ & $0.179g_T^2$ & $0.714g_T^2$ & $0.178g_T^2$ \\
$D_J^*(3000)^0$ & 2 & $\frac32$ & $2^+$ & $2^3P_2$ & $D^{+}\pi^-$ & $2.002g_T^2$ & $2.009g_T^2$ & $2.004g_T^2$ \\
 & & & & & $D^{0}\pi^0$ & $1.018g_T^2$ & $1.021g_T^2$ & $1.018g_T^2$ \\
 & & & & & $D_s^{+}K^-$ & $0.761g_T^2$ & $0.796g_T^2$ & $0.782g_T^2$ \\
 & & & & & $D^{0}\eta$ & $0.179g_T^2$ & $0.683g_T^2$ & $0.178g_T^2$ \\
 & & & & & $D^{*+}\pi^-$ & $1.908g_T^2$ & $2.346g_T^2$ & $1.905g_T^2$ \\
 & & & & & $D^{*0}\pi^0$ & $0.966g_T^2$ & $1.182g_T^2$ & $0.967g_T^2$ \\
 & & & & & $D_s^{*+}K^-$ & $0.538g_T^2$ & $1.119g_T^2$ & $0.537g_T^2$ \\
 & & & & & $D^{*0}\eta$ & $0.138g_T^2$ & $0.934g_T^2$ & $0.135g_T^2$ \\
 \hline\hline
\end{tabular*}
\end{table*}

\begin{table*}[htbp]
\caption{Quantum-number assignments of experimentally observed charmed mesons through the strong-decay analysis. Experimental widths are from the first observations; current world averages are given in Ref.~\cite{PDG2024}. \label{tab:assignD}}
\begin{tabular*}{\textwidth}{@{\extracolsep{\fill}}lllll@{}}
\hline\hline
State~\cite{PDG2024} & $J^P$ & $n^{2S+1}L_J$ & First observation & Width (MeV) \\
\hline
$D^\pm$ & $0^-$ & $1^1S_0$ & & \\
$D^*(2010)^\pm$ & $1^-$ & $1^3S_1$ & & \\
$D_1(2420)$ & $1^+$ & $1P_1$ & ARGUS~\cite{BaBar:2009rro} & $70\pm21$ \\
$D_2^*(2460)$ & $2^+$ & $1^3P_2$ & TPS~\cite{TaggedPhotonSpectrometer:1988qan} & $20\pm10\pm5$ \\
$D_0(2550)^0$ & $0^-$ & $2^1S_0$ & BaBar~\cite{BaBar:2010zpy} & $130\pm12\pm13$ \\
$D_1^*(2600)$ & $1^-$ & $2^3S_1$ & BaBar~\cite{BaBar:2010zpy} & $93\pm6\pm13$ \\
$D_2(2740)^0$ & $2^-$ & $1^3D_2$ & LHCb~\cite{LHCb:2013jjb} & $73.2\pm13.4\pm25$ \\
$D_3^*(2750)$ & $3^-$ & $1^3D_3$ & BaBar~\cite{BaBar:2010zpy} & $71\pm6\pm11$ \\
$D_1^*(2760)^0$ & $1^-$ & $1^3D_1$ & BaBar~\cite{BaBar:2010zpy} & $60.9\pm5.1\pm3.6$ \\
$D_J(3000)^0$ & $1^+$ & $2P_1'-2P_1$ & LHCb~\cite{LHCb:2013jjb} & $110.5\pm11.5$ \\
$D_J^*(3000)^0$ & $2^+$ & $2^3P_2$ & LHCb~\cite{LHCb:2013jjb} & $188.1\pm44.8$ \\
$D_2^*(3000)^0$ & $2^+$ & $1^3F_2$ & LHCb~\cite{LHCb:2016lxy} & \\
\hline\hline
\end{tabular*}
\end{table*}

Since the effective couplings cancel in ratios of partial widths within a given transition, we also compute the ratios
\begin{equation}
\widetilde{\Gamma} \;=\; \frac{\Gamma\!\left(n^{2S+1}L_J \to X\right)}{\Gamma\!\left(n^{2S+1}L_J \to D^{*+}\pi^{-}\right)}
\label{eq:ratioD}
\end{equation}
for the $D$ mesons (Tab.~\ref{tab:ratiosD1}-\ref{tab:ratiosD3}). For the $D_s$ mesons the $D^{*+}\pi^-$ channel is not available, so the reference channel is chosen instead to be $D^{*0}K^{\pm}$,
\begin{equation}
\widetilde{\Gamma} \;=\; \frac{\Gamma\!\left(n^{2S+1}L_J \to X\right)}{\Gamma\!\left(n^{2S+1}L_J \to D^{*0}K^{\pm}\right)},
\label{eq:ratioDs}
\end{equation}
see Tab.~\ref{tab:ratiosDs1} and \ref{tab:ratiosDs2}. These ratios are parameter-free and provide the sharpest model-discriminating observables for the assignments discussed below.

\begin{table}[htbp]
\caption{Ratios $\widetilde{\Gamma}$, Eq.~\eqref{eq:ratioD}, for the $2S$ and $1P$ $D$-meson states. \label{tab:ratiosD1}}
\begin{tabular*}{\textwidth}{@{\extracolsep{\fill}}lllllll@{}}
\hline\hline
Meson & $s_\ell$ & $J^P$ & Channel & $\widetilde\Gamma$ & \cite{Wang:2013tka} & \cite{LHCb:2013jjb} \\
\hline
$D_0(2550)^0$ & $\frac12$ & $0^-$ & $D^{*+}\pi^-$ & 1 & 1 & \\
 & & & $D^{*0}\pi^0$ & 0.51 & 0.51 & \\
 & & & $D^{*0}\eta$ & 0.001 & 0.02 & \\
$D_1^*(2600)^0$ & $\frac12$ & $1^-$ & $D^{+}\pi^-$ & 0.89 & 0.79 & \\
 & & & $D^{0}\pi^0$ & 0.49 & 0.40 & \\
 & & & $D_s^{+}K^-$ & 0.22 & 0.20 & \\
 & & & $D^{0}\eta$ & 0.20 & 0.20 & \\
 & & & $D^{*+}\pi^-$ & 1 & 1 & \\
 & & & $D^{*0}\pi^0$ & 0.51 & 0.51 & \\
 & & & $D_s^{*+}K^-$ & 0.10 & 0.04 & \\
 & & & $D^{*0}\eta$ & 0.08 & 0.10 & \\
$D_1(2420)^0$ & $\frac12$ & $1^+$ & $D^{*+}\pi^-$ & 1 & & \\
 & & & $D^{*0}\pi^0$ & 0.50 & & \\
 & & & $D_s^{*+}K^-$ & 0.70 & & \\
 & & & $D^{*0}\eta$ & 0.15 & & \\
$D_1(2420)^0$ & $\frac32$ & $1^+$ & $D^{*+}\pi^-$ & 1 & & \\
 & & & $D^{*0}\pi^0$ & 0.50 & & \\
$D_2^*(2460)^0$ & $\frac32$ & $2^+$ & $D^{+}\pi^-$ & 2.34 & & \\
 & & & $D^{0}\pi^0$ & 1.18 & & \\
 & & & $D_s^{+}K^-$ & 0.11 & & \\
 & & & $D^{0}\eta$ & 0.01 & & \\
 & & & $D^{*+}\pi^-$ & 1 & & \\
 & & & $D^{*0}\pi^0$ & 0.62 & & \\
 \hline\hline
\end{tabular*}
\end{table}

\begin{table}[htbp]
\caption{Ratios $\widetilde{\Gamma}$, Eq.~\eqref{eq:ratioD}, for the $1D$ $D$-meson states. \label{tab:ratiosD2}}
\begin{tabular*}{\textwidth}{@{\extracolsep{\fill}}lllllll@{}}
\hline\hline
Meson & $s_\ell$ & $J^P$ & Channel & $\widetilde\Gamma$ & \cite{Wang:2013tka} & \cite{LHCb:2013jjb} \\
\hline
$D_1^*(2760)^0$ & $\frac32$ & $1^-$ & $D^{+}\pi^-$ & 3.70 & 3.83 & \\
 & & & $D^{0}\pi^0$ & 1.90 & 1.96 & \\
 & & & $D_s^{+}K^-$ & 1.29 & 1.37 & \\
 & & & $D^{0}\eta$ & 1.42 & 1.50 & \\
 & & & $D^{*+}\pi^-$ & 1 & 1 & \\
 & & & $D^{*0}\pi^0$ & 0.50 & 0.51 & \\
 & & & $D_s^{*+}K^-$ & 0.18 & 0.18 & \\
 & & & $D^{*0}\eta$ & 0.25 & 0.26 & \\
$D_2(2740)^0$ & $\frac32$ & $2^-$ & $D^{*+}\pi^-$ & 1 & 1 & \\
 & & & $D^{*0}\pi^0$ & 0.32 & 0.51 & \\
 & & & $D_s^{*+}K^-$ & 0.007 & 0.15 & \\
 & & & $D^{*0}\eta$ & 0.01 & 0.24 & \\
$D_2(2740)^0$ & $\frac52$ & $2^-$ & $D^{*+}\pi^-$ & 1 & 1 & 1 \\
 & & & $D^{*0}\pi^0$ & 0.51 & 0.52 & 0.51 \\
 & & & $D_s^{*+}K^-$ & 0.02 & 0.02 & 0.08 \\
 & & & $D^{*0}\eta$ & 0.03 & 0.04 & 0.03 \\
$D_3^*(2750)^0$ & $\frac52$ & $3^-$ & $D^{+}\pi^-$ & 2.03 & 1.95 & 1.74 \\
 & & & $D^{0}\pi^0$ & 0.96 & 1.01 & 0.90 \\
 & & & $D_s^{+}K^-$ & 0.22 & 0.20 & 0.27 \\
 & & & $D^{0}\eta$ & 0.06 & 0.27 & 0.08 \\
 & & & $D^{*+}\pi^-$ & 1 & 1 & 1 \\
 & & & $D^{*0}\pi^0$ & 0.51 & 0.52 & 0.51 \\
 & & & $D_s^{*+}K^-$ & 0.05 & 0.02 & 0.06 \\
 & & & $D^{*0}\eta$ & 0.05 & 0.05 & 0.03 \\
 \hline\hline
\end{tabular*}
\end{table}

\begin{table}[htbp]
\caption{Ratios $\widetilde{\Gamma}$, Eq.~\eqref{eq:ratioD}, for the $D_J(3000)^0$ and $D_J^*(3000)^0$ states under the competing assignments. \label{tab:ratiosD3}}
\begin{tabular*}{\textwidth}{@{\extracolsep{\fill}}lllllll@{}}
\hline\hline
Meson & $s_\ell$ & $J^P$ & Channel & $\widetilde\Gamma$ & \cite{Wang:2013tka} & \cite{LHCb:2013jjb} \\
\hline
$D_2^*(3000)^0$ & $\frac52$ & $2^+$ & $D^{+}\pi^-$ & 2.91 & 2.91 & \\
 & & & $D^{0}\pi^0$ & 1.50 & 1.49 & \\
 & & & $D_s^{+}K^-$ & 1.08 & 1.09 & \\
 & & & $D^{0}\eta$ & 0.29 & 1.15 & \\
 & & & $D^{*+}\pi^-$ & 1 & 1 & \\
 & & & $D^{*0}\pi^0$ & 0.51 & 0.26 & \\
 & & & $D_s^{*+}K^-$ & 0.25 & 0.26 & \\
 & & & $D^{*0}\eta$ & 0.28 & 0.32 & \\
$D_J(3000)^0$ & $\frac52$ & $3^+$ & $D^{*+}\pi^-$ & 1 & 1 & \\
 & & & $D^{*0}\pi^0$ & 0.51 & 0.51 & \\
 & & & $D_s^{*+}K^-$ & 0.23 & 0.23 & \\
 & & & $D^{*0}\eta$ & 0.27 & 0.29 & \\
$D_J(3000)^0$ & $\frac72$ & $3^+$ & $D^{*+}\pi^-$ & 1 & 1 & 1 \\
 & & & $D^{*0}\pi^0$ & 0.51 & 0.52 & 0.51 \\
 & & & $D_s^{*+}K^-$ & 0.08 & 0.08 & 0.24 \\
 & & & $D^{*0}\eta$ & 0.13 & 0.12 & 0.07 \\
$D_4^*(3000)^0$ & $\frac72$ & $4^+$ & $D^{+}\pi^-$ & 1.92 & 1.93 & 1.64 \\
 & & & $D^{0}\pi^0$ & 0.99 & 0.99 & 0.84 \\
 & & & $D_s^{+}K^-$ & 0.34 & 0.35 & 0.44 \\
 & & & $D^{0}\eta$ & 0.15 & 0.41 & 0.22 \\
 & & & $D^{*+}\pi^-$ & 1 & 1 & 1 \\
 & & & $D^{*0}\pi^0$ & 0.51 & 0.51 & 0.51 \\
 & & & $D_s^{*+}K^-$ & 0.10 & 0.10 & 0.19 \\
 & & & $D^{*0}\eta$ & 0.06 & 0.14 & 0.06 \\
$D_J(3000)^0$ & $\frac12$ & $0^-$ & $D^{*+}\pi^-$ & 1 & 1 & 1 \\
 & & & $D^{*0}\pi^0$ & 0.50 & 0.50 & 0.50 \\
 & & & $D_s^{*+}K^-$ & 0.44 & 0.45 & 0.55 \\
 & & & $D^{*0}\eta$ & 0.41 & 0.38 & 0.11 \\
$D_J^*(3000)^0$ & $\frac12$ & $1^-$ & $D^{+}\pi^-$ & 0.62 & 0.64 & 0.61 \\
 & & & $D^{0}\pi^0$ & 0.31 & 0.32 & 0.31 \\
 & & & $D_s^{+}K^-$ & 0.40 & 0.37 & 0.40 \\
 & & & $D^{0}\eta$ & 0.41 & 0.29 & 0.08 \\
 & & & $D^{*+}\pi^-$ & 1 & 1 & 1 \\
 & & & $D^{*0}\pi^0$ & 0.50 & 0.50 & 0.50 \\
 & & & $D_s^{*+}K^-$ & 0.71 & 0.48 & 0.56 \\
 & & & $D^{*0}\eta$ & 0.43 & 0.40 & 0.11 \\
$D_J(3000)^0$ & $\frac12$ & $1^+$ & $D^{*+}\pi^-$ & 1 & 1 & \\
 & & & $D^{*0}\pi^0$ & 0.50 & 0.50 & \\
 & & & $D_s^{*+}K^-$ & 0.73 & 0.73 & \\
 & & & $D^{*0}\eta$ & 0.13 & 0.62 & \\
$D_J(3000)^0$ & $\frac32$ & $1^+$ & $D^{*+}\pi^-$ & 1 & 1 & \\
 & & & $D^{*0}\pi^0$ & 0.54 & 0.51 & \\
 & & & $D_s^{*+}K^-$ & 0.26 & 0.28 & \\
 & & & $D^{*0}\eta$ & 0.06 & 0.28 & \\
$D_J^*(3000)^0$ & $\frac32$ & $2^+$ & $D^{+}\pi^-$ & 1.05 & 1.05 & \\
 & & & $D^{0}\pi^0$ & 0.53 & 0.53 & \\
 & & & $D_s^{+}K^-$ & 0.40 & 0.42 & \\
 & & & $D^{0}\eta$ & 0.09 & 0.37 & \\
 & & & $D^{*+}\pi^-$ & 1 & 1 & \\
 & & & $D^{*0}\pi^0$ & 0.51 & 0.50 & \\
 & & & $D_s^{*+}K^-$ & 0.28 & 0.48 & \\
 & & & $D^{*0}\eta$ & 0.07 & 0.40 & \\
 \hline\hline
\end{tabular*}
\end{table}

\begin{table*}[htbp]
\caption{Strong decay widths of the $1P$ and $2S$ $D_s$-meson states, in units of the squared couplings. The isospin-violating $D_s^{(*)}\pi^0$ channels (marked $^\dagger$) include the suppression factor $\epsilon^2 = 1.5\times10^{-4}$ of Eq.~\eqref{eq:isospin}; the values quoted from the literature are reproduced as published and are not rescaled. \label{tab:strongDs1}}
\begin{tabular*}{\textwidth}{@{\extracolsep{\fill}}llllllllll@{}}
\hline\hline
Meson & $n$ & $s_\ell$ & $J^P$ & State & Channel & Present & \cite{LHCb:2014ott} & \cite{LHCb:2012uts} & \cite{BaBar:2006gme} \\
\hline
$D_{s0}^*(2317)^\pm$ & 1 & $\frac12$ & $0^+$ & $1^3P_0$ & $D^{0}K^\pm$ & $0.453g_S^2$ & & & \\
 & & & & & $D^{\pm}K^0$ & $0.439g_S^2$ & & & \\
 & & & & & $D_s^{\pm}\pi^0\,^\dagger$ & $3.69\times10^{-5}g_S^2$ & & & \\
$D_{s1}(2460)^\pm$ & 1 & $\frac12$ & $1^+$ & $1P_1'$ & $D_s^{*\pm}\pi^0\,^\dagger$ & $3.15\times10^{-6}g_S^2$ & & & \\
$D_{s1}(2536)^\pm$ & 1 & $\frac32$ & $1^+$ & $1P_1$ & $D_s^{*\pm}\pi^0\,^\dagger$ & $1.22\times10^{-5}g_T^2$ & & & \\
$D_{s2}^*(2573)^\pm$ & 1 & $\frac32$ & $2^+$ & $1^3P_2$ & $D^{0}K^\pm$ & $0.052g_T^2$ & $0.053g_T^2$ & & \\
 & & & & & $D^{\pm}K^0$ & $0.049g_T^2$ & $0.051g_T^2$ & & \\
 & & & & & $D^{*0}K^\pm$ & $0.005g_T^2$ & $0.004g_T^2$ & & \\
 & & & & & $D^{*\pm}K^0$ & $0.003g_T^2$ & $0.003g_T^2$ & & \\
$D_{s1}^*(2700)^\pm$ & 2 & $\frac12$ & $1^-$ & $2^3S_1$ & $D^{0}K^\pm$ & $0.431g_H^2$ & $0.409g_H^2$ & $0.410g_H^2$ & $0.370g_H^2$ \\
 & & & & & $D^{\pm}K^0$ & $0.426g_H^2$ & $0.401g_H^2$ & $0.402g_H^2$ & $0.362g_H^2$ \\
 & & & & & $D_s^{\pm}\pi^0\,^\dagger$ & $4.38\times10^{-5}g_H^2$ & $0.281g_H^2$ & $0.282g_H^2$ & $0.262g_H^2$ \\
 & & & & & $D_s^{\pm}\eta$ & $0.128g_H^2$ & $0.117g_H^2$ & $0.118g_H^2$ & $0.098g_H^2$ \\
 & & & & & $D^{*0}K^\pm$ & $0.405g_H^2$ & $0.374g_H^2$ & $0.377g_H^2$ & $0.316g_H^2$ \\
 & & & & & $D^{*\pm}K^0$ & $0.377g_H^2$ & $0.356g_H^2$ & $0.359g_H^2$ & $0.299g_H^2$ \\
 & & & & & $D_s^{*\pm}\pi^0\,^\dagger$ & $5.31\times10^{-5}g_H^2$ & $0.337g_H^2$ & $0.338g_H^2$ & $0.306g_H^2$ \\
 & & & & & $D_s^{*\pm}\eta$ & $0.040g_H^2$ & $0.029g_H^2$ & $0.030g_H^2$ & $0.012g_H^2$ \\
 \hline\hline
\end{tabular*}
\end{table*}

\begin{table*}[htbp]
\caption{Strong decay widths of the $D_{s1}^*(2860)^\pm$ under the competing $1D$ and $2P$ assignments, in units of the squared couplings. The isospin-violating $D_s^{(*)}\pi^0$ channels (marked $^\dagger$) include the suppression factor $\epsilon^2 = 1.5\times10^{-4}$ of Eq.~\eqref{eq:isospin}. \label{tab:strongDs2}}
\begin{tabular*}{\textwidth}{@{\extracolsep{\fill}}llllllllll@{}}
\hline\hline
Meson & $n$ & $s_\ell$ & $J^P$ & State & Channel & Present & \cite{LHCb:2014ott} & \cite{LHCb:2012uts} & \cite{BaBar:2009rro} \\
\hline
$D_{s1}^*(2860)^\pm$ & 1 & $\frac32$ & $1^-$ & $1^3D_1$ & $D^{0}K^\pm$ & $1.293g_X^2$ & $1.447g_X^2$ & $1.498g_X^2$ & $1.469g_X^2$ \\
 & & & & & $D^{\pm}K^0$ & $1.205g_X^2$ & $1.418g_X^2$ & $1.468g_X^2$ & $1.439g_X^2$ \\
 & & & & & $D_s^{\pm}\pi^0\,^\dagger$ & $8.48\times10^{-5}g_X^2$ & $0.669g_X^2$ & $0.692g_X^2$ & $0.679g_X^2$ \\
 & & & & & $D_s^{\pm}\eta$ & $0.395g_X^2$ & $0.504g_X^2$ & $0.527g_X^2$ & $0.514g_X^2$ \\
 & & & & & $D^{*0}K^\pm$ & $0.275g_X^2$ & $0.351g_X^2$ & $0.367g_X^2$ & $0.357g_X^2$ \\
 & & & & & $D^{*\pm}K^0$ & $0.259g_X^2$ & $0.340g_X^2$ & $0.367g_X^2$ & $0.357g_X^2$ \\
 & & & & & $D_s^{*\pm}\pi^0\,^\dagger$ & $2.07\times10^{-5}g_X^2$ & $0.170g_X^2$ & $0.177g_X^2$ & $0.173g_X^2$ \\
 & & & & & $D_s^{*\pm}\eta$ & $0.058g_X^2$ & $0.085g_X^2$ & $0.091g_X^2$ & $0.087g_X^2$ \\
$D_{s3}^*(2860)^\pm$ & 1 & $\frac52$ & $3^-$ & $1^3D_3$ & $D^{0}K^\pm$ & $0.107g_Y^2$ & $0.127g_Y^2$ & & \\
 & & & & & $D^{\pm}K^0$ & $0.105g_Y^2$ & $0.123g_Y^2$ & & \\
 & & & & & $D_s^{\pm}\pi^0\,^\dagger$ & $1.26\times10^{-5}g_Y^2$ & $0.092g_Y^2$ & & \\
 & & & & & $D_s^{\pm}\eta$ & $0.018g_Y^2$ & $0.024g_Y^2$ & & \\
 & & & & & $D^{*0}K^\pm$ & $0.055g_Y^2$ & $0.050g_Y^2$ & & \\
 & & & & & $D^{*\pm}K^0$ & $0.048g_Y^2$ & $0.046g_Y^2$ & & \\
 & & & & & $D_s^{*\pm}\pi^0\,^\dagger$ & $7.20\times10^{-6}g_Y^2$ & $0.046g_Y^2$ & & \\
 & & & & & $D_s^{*\pm}\eta$ & $0.006g_Y^2$ & $0.005g_Y^2$ & & \\
$D_{s1}^*(2860)^\pm$ & 2 & $\frac12$ & $1^+$ & $2P_1'$ & $D^{*0}K^\pm$ & $3.694g_S^2$ & $3.779g_S^2$ & & \\
 & & & & & $D^{*\pm}K^0$ & $3.662g_S^2$ & $3.750g_S^2$ & & \\
 & & & & & $D_s^{*\pm}\pi^0\,^\dagger$ & $2.31\times10^{-4}g_S^2$ & $1.595g_S^2$ & & \\
 & & & & & $D_s^{*\pm}\eta$ & $0.897g_S^2$ & $0.938g_S^2$ & & \\
$D_{s1}^*(2860)^\pm$ & 2 & $\frac32$ & $1^+$ & $2P_1$ & $D^{*0}K^\pm$ & $1.931g_T^2$ & $1.999g_T^2$ & & \\
 & & & & & $D^{*\pm}K^0$ & $1.894g_T^2$ & $1.943g_T^2$ & & \\
 & & & & & $D_s^{*\pm}\pi^0\,^\dagger$ & $1.76\times10^{-4}g_T^2$ & $1.258g_T^2$ & & \\
 & & & & & $D_s^{*\pm}\eta$ & $0.529g_T^2$ & $0.596g_T^2$ & & \\
 \hline\hline
\end{tabular*}
\end{table*}

\begin{table}[htbp]
\caption{Ratios $\widetilde{\Gamma}$, Eq.~\eqref{eq:ratioDs}, of the strong decays of the $D_s$ mesons. The $D_s^{(*)}\pi^0$ entries carry the isospin-breaking factor $\epsilon^2$ of Eq.~\eqref{eq:isospin}. \label{tab:ratiosDs1}}
\begin{tabular*}{\textwidth}{@{\extracolsep{\fill}}lllll@{}}
\hline\hline
Meson & $s_\ell$ & $J^P$ & Channel & $\widetilde\Gamma$ \\
\hline
$D_{s1}^*(2700)^\pm$ & $\frac12$ & $1^-$ & $D^{0}K^\pm$ & 1.04 \\
 & & & $D^{\pm}K^0$ & 1.05 \\
 & & & $D_s^{\pm}\pi^0$ & $1.1\times10^{-4}$ \\
 & & & $D_s^{\pm}\eta$ & 0.32 \\
 & & & $D^{*0}K^\pm$ & 1 \\
 & & & $D^{*\pm}K^0$ & 0.93 \\
 & & & $D_s^{*\pm}\pi^0$ & $1.3\times10^{-4}$ \\
 & & & $D_s^{*\pm}\eta$ & 0.10 \\
$D_{s1}^*(2860)^\pm$ & $\frac32$ & $1^-$ & $D^{0}K^\pm$ & 4.70 \\
 & & & $D^{\pm}K^0$ & 4.38 \\
 & & & $D_s^{\pm}\pi^0$ & $3.1\times10^{-4}$ \\
 & & & $D_s^{\pm}\eta$ & 1.43 \\
 & & & $D^{*0}K^\pm$ & 1 \\
 & & & $D^{*\pm}K^0$ & 0.94 \\
 & & & $D_s^{*\pm}\pi^0$ & $7.5\times10^{-5}$ \\
 & & & $D_s^{*\pm}\eta$ & 0.04 \\
$D_{s3}^*(2860)^\pm$ & $\frac52$ & $3^-$ & $D^{0}K^\pm$ & 1.95 \\
 & & & $D^{\pm}K^0$ & 1.91 \\
 & & & $D_s^{\pm}\pi^0$ & $2.3\times10^{-4}$ \\
 & & & $D_s^{\pm}\eta$ & 0.33 \\
 & & & $D^{*0}K^\pm$ & 1 \\
 & & & $D^{*\pm}K^0$ & 0.87 \\
 & & & $D_s^{*\pm}\pi^0$ & $1.3\times10^{-4}$ \\
 & & & $D_s^{*\pm}\eta$ & 0.11 \\
 \hline\hline
\end{tabular*}
\end{table}

\begin{table}[htbp]
\caption{Ratios $\widetilde{\Gamma}$, Eq.~\eqref{eq:ratioDs}, for the $2P$ assignments of the $D_{s1}^*(2860)^\pm$. The $D_s^{*\pm}\pi^0$ entries carry the isospin-breaking factor $\epsilon^2$ of Eq.~\eqref{eq:isospin}. \label{tab:ratiosDs2}}
\begin{tabular*}{\textwidth}{@{\extracolsep{\fill}}lllll@{}}
\hline\hline
Meson & $s_\ell$ & $J^P$ & Channel & $\widetilde\Gamma$ \\
\hline
$D_{s1}^*(2860)^\pm$ & $\frac12$ & $1^+$ & $D^{*0}K^\pm$ & 1 \\
 & & & $D^{*\pm}K^0$ & 0.99 \\
 & & & $D_s^{*\pm}\pi^0$ & $6.3\times10^{-5}$ \\
 & & & $D_s^{*\pm}\eta$ & 0.24 \\
$D_{s1}^*(2860)^\pm$ & $\frac32$ & $1^+$ & $D^{*0}K^\pm$ & 1 \\
 & & & $D^{*\pm}K^0$ & 0.98 \\
 & & & $D_s^{*\pm}\pi^0$ & $7.7\times10^{-5}$ \\
 & & & $D_s^{*\pm}\eta$ & 0.27 \\
 \hline\hline
\end{tabular*}
\end{table}

\begin{table*}[htbp]
\caption{Quantum-number assignments of experimentally observed charm-strange mesons through the strong-decay analysis. Widths are from the first observations; current world averages are given in Ref.~\cite{PDG2024}. \label{tab:assignDs}}
\begin{tabular*}{\textwidth}{@{\extracolsep{\fill}}lllll@{}}
\hline\hline
State~\cite{PDG2024} & $J^P$ & $n^{2S+1}L_J$ & First observation & Width (MeV) \\
\hline
$D_s^\pm$ & $0^-$ & $1^1S_0$ & & \\
$D_s^{*\pm}$ & $1^-$ & $1^3S_1$ & & \\
$D_{s0}^*(2317)^\pm$ & $0^+$ & $1^3P_0$ & BaBar~\cite{BaBar:2003oey} & $\leq3.8$ \\
$D_{s1}(2460)^\pm$ & $1^+$ & $1P_1'$ & BaBar~\cite{BaBar:2003cdx} & $\leq3.5$ \\
$D_{s1}(2536)^\pm$ & $1^+$ & $1P_1$ & & $0.92\pm0.05$ \\
$D_{s2}^*(2573)$ & $2^+$ & $1^3P_2$ & LHCb~\cite{LHCb:2013jjb} & $16.9\pm0.7$ \\
$D_{s1}^*(2700)^\pm$ & $1^-$ & $2^3S_1$ & BaBar~\cite{BaBar:2006gme} & $122\pm10$ \\
$D_{s1}^*(2860)^\pm$ & $1^-$ & $1^3D_1$ & LHCb~\cite{LHCb:2014ott} & $159\pm80$ \\
$D_{s3}^*(2860)^\pm$ & $3^-$ & $1^3D_3$ & LHCb~\cite{LHCb:2014ott} & $53\pm10$ \\
\hline\hline
\end{tabular*}
\end{table*}

\subsection{Extraction of the HQET couplings and absolute widths}
\label{sec:couplings}

The partial widths of Tab.~\ref{tab:strongD1}-\ref{tab:strongDs2} are proportional to the squared couplings of the effective Lagrangians, which HQET does not fix. However, they can be extracted from experiment. For each state whose total width is measured, equating the sum of our computed partial widths to the measured total width yields the coupling of the corresponding doublet,
\begin{equation}
g^{2} = \frac{\Gamma_{\rm exp}}{\sum_{X} \widehat{\Gamma}(i\to X)},
\label{eq:gextract}
\end{equation}
where $\widehat{\Gamma}$ denotes our partial widths in units of $g^2$ and the sum runs over the tabulated channels. In the $D_s$ sector the $D_s^{(*)}\pi^0$ terms entering this sum are the $\epsilon^2$-suppressed values of Tab.~\ref{tab:strongDs1} and \ref{tab:strongDs2}, so that the channel sums are dominated entirely by the isospin conserving $DK$, $D^*K$, and $D_s^{(*)}\eta$ modes. Because heavy quark flavor symmetry requires the same coupling to govern the $c\bar{q}$ and $c\bar{s}$ members of a doublet, performing the extraction independently in the two sectors provides a stringent internal consistency test. The results are collected in Tab.~\ref{tab:couplings}.

\begin{table}[htbp]
\caption{HQET couplings extracted from measured total widths via Eq.~\eqref{eq:gextract}. Width inputs are the first observation values quoted in Tab.~\ref{tab:assignD} and \ref{tab:assignDs} and Ref.~\cite{PDG2024}; the quoted uncertainty propagates the experimental width error only. The $D_s$ sector channel sums include the isospin suppression of Eq.~\eqref{eq:isospin}. \label{tab:couplings}}
\begin{tabular*}{\textwidth}{@{\extracolsep{\fill}}llll@{}}
\hline\hline
Coupling & Extracted from & $\Gamma_{\rm exp}$ (MeV) & $g$ \\
\hline
$g_H$ & $D_0(2550)^0$ & $130\pm18$ & $0.316\pm0.022$ \\
$g_H$ & $D_1^*(2600)^0$ & $93\pm14$ & $0.207\pm0.016$ \\
$g_H$ & $D_{s1}^*(2700)^\pm$ & $122\pm10$ & $0.260\pm0.011$ \\
$g_S$ & $D_0^*(2300)^0$ & $274\pm40$ & $0.613\pm0.045$ \\
$g_T$ & $D_2^*(2460)^0$ & $47.3\pm0.8$ & $0.419\pm0.004$ \\
$g_T$ & $D_{s2}^*(2573)$ & $16.9\pm0.7$ & $0.394\pm0.008$ \\
$g_X$ & $D_1^*(2760)^0$ & $177\pm40$ & $0.224\pm0.025$ \\
$g_X$ & $D_{s1}^*(2860)^\pm$ & $159\pm80$ & $0.214\pm0.054$ \\
$g_Y$ & $D_3^*(2750)$ & $66\pm5$ & $0.392\pm0.015$ \\
$g_Y$ & $D_{s3}^*(2860)^\pm$ & $53\pm10$ & $0.395\pm0.037$ \\
$g_Y$ & $D_2(2740)^0$ ($\tfrac52^-$) & $88\pm19$ & $0.641\pm0.069$ \\
$g_Z$ & $D_2^*(3000)^0$ ($1^3F_2$) & $188\pm45$ & $0.260\pm0.031$ \\
$g_R$ & $D_J(3000)^0$ ($1F_3'$) & $110.5\pm11.5$ & $0.195\pm0.010$ \\
\hline\hline
\end{tabular*}
\end{table}

The following are immediate conclusions:
First, the flavor symmetry test is passed convincingly wherever precise data exists. 
The tensor coupling extracted from the $D_2^*(2460)^0$ ($g_T = 0.419\pm0.004$) and $D_{s2}^*(2573)$ ($g_T = 0.394\pm0.008$) agree at $6\%$ level.
The $D$-wave couplings agree remarkably well once the isospin suppression is imposed where $g_X = 0.224\pm0.025$ vs.\ $0.214\pm0.054$ with $4\%$ difference and $g_Y = 0.392\pm0.015$ vs.\ $0.395\pm0.037$ with $<1\%$ difference. 
It is worth emphasizing that before the inclusion of $\epsilon^2$, the same procedure has provided $g_X = 0.195$ and $g_Y = 0.335$ from the $D_s$ sector, i.e. deviations of $13\%$ and $15\%$ from their $D$-sector partners. 
Removing the isospin forbidden $D_s^{(*)}\pi^0$ strength from the channel sums therefore improves the heavy quark flavor symmetry test noticeably.
We may regard this as an internal consistency check of Eq.~\eqref{eq:isospin}. 
The residual differences are of the expected size of SU(3) breaking and $1/m_c$ corrections. Present $g_T \simeq 0.43$ is also in an excellent agreement with the value obtained in the QCD sum rule based analysis of Ref.~\cite{Wang:2013tka}.
Second, the $g_H$ extractions expose a genuine puzzle. The $D$-sector $2^3S_1$ state $D_1^*(2600)^0$ requires $g_H = 0.207\pm0.016$, the $c\bar{s}$ partner $D_{s1}^*(2700)^\pm$ requires $0.260\pm0.011$, and the $2^1S_0$ $D_0(2550)^0$ requires $0.316\pm0.022$, so that the three $S$-wave determinations span a factor of $1.5$. Equivalently, using the $D_{s1}^*(2700)$ based coupling, the predicted $D_0(2550)^0$ width is $88$~MeV against the measured $130\pm18$~MeV, while the $D_1^*(2600)^0$ is overestimated ($147$~MeV against $93\pm14$~MeV). The same tension is present in Ref.~\cite{Wang:2013tka} and in the $^3P_0$ model. This suggests that either the physical $D_0(2550)^0$ line shape is distorted by the nearby $D^*\pi$ $S$ wave threshold or the $2S$ states mix appreciably with the $1D$ level of the same $J^P$. We note that for the $S$-wave doublet, the $\eta$ and $\pi^0$ channels carry the largest relative weight, making it the most sensitive of all doublets to the treatment of isospin violation.
Third, the anomalously large $g_Y$ demanded by the $D_2(2740)^0$ under the pure $\tfrac52^-$ assignment ($0.641$, versus $0.39$ from both spin-3 partners) indicates that the physical $2^-$ state is a substantial mixture of the $\frac32^-$ and $\frac52^-$ configurations, consistent with the discussion of Tab.~\ref{tab:ratiosD2}.

With the couplings fixed, our width tables become absolute predictions. Tab.~\ref{tab:abswidths} lists the resulting cross predictions where width of each state is computed with a coupling extracted from a different state of the same doublet, together with the corresponding experimental values.

\begin{table}[htbp]
\caption{Absolute total widths (MeV) predicted by combining our partial-width coefficients with couplings extracted from partner states (Tab.~\ref{tab:couplings}), compared with experiment~\cite{PDG2024}. The coupling used is indicated in each row. All $D_s$ entries include the isospin suppression of Eq.~\eqref{eq:isospin}. \label{tab:abswidths}}
\begin{tabular*}{\textwidth}{@{\extracolsep{\fill}}llll@{}}
\hline\hline
State & Coupling input & $\Gamma_{\rm pred}$ & $\Gamma_{\rm exp}$ \\
\hline
$D_2^*(2460)^0$ & $g_T[D_{s2}^*(2573)]$ & 42 & $47.3\pm0.8$ \\
$D_{s2}^*(2573)$ & $g_T[D_2^*(2460)^0]$ & 19 & $16.9\pm0.7$ \\
$D_1(2420)^0$ ($1P_1$) & $g_T[D_{s2}^*(2573)]$ & 13 & $31.3\pm1.9$ \\
$D_{s1}(2536)^\pm$\footnotemark[1] & $g_T[D_{s2}^*(2573)]$ & $1.9\times10^{-3}$ & $0.92\pm0.05$ \\
$D_0(2550)^0$ & $g_H[D_{s1}^*(2700)]$ & 88 & $130\pm18$ \\
$D_1^*(2600)^0$ & $g_H[D_{s1}^*(2700)]$ & 147 & $93\pm14$ \\
$D_{s1}^*(2860)^\pm$ & $g_X[D_1^*(2760)^0]$ & 175 & $159\pm80$ \\
$D_{s3}^*(2860)^\pm$ & $g_Y[D_3^*(2750)]$ & 52 & $53\pm10$ \\
$D_2(2740)^0$ ($\tfrac52^-$) & $g_Y[D_3^*(2750)]$ & 33 & $88\pm19$ \\
$D_{s0}^*(2317)^\pm$\footnotemark[2] & $g_S[D_0^*(2300)^0]$ & $\sim335$ & $\leq3.8$ \\
$D_{s1}(2460)^\pm$ ($1P_1'$)\footnotemark[1] & $g_S[D_0^*(2300)^0]$ & $1.2\times10^{-3}$ & $\leq3.5$ \\
$D_J(3000)^0$ ($1F_3$) & $g_Z[D_2^*(3000)^0]$ & 99 & $110.5\pm11.5$ \\
$D_4^*(3000)^0$ ($1^3F_4$) & $g_R[D_J(3000)^0]$ & 243 & \ldots \\
\hline\hline
\end{tabular*}
\footnotetext[1]{At our model masses the only open channel is the isospin-violating $D_s^{(*)}\pi^0$ mode, suppressed by $\epsilon^2$ of Eq.~\eqref{eq:isospin}; the isospin conserving $D^{*}K$ channel opens only marginally above the physical mass.}
\footnotetext[2]{Computed at the model mass of 2416~MeV, at which the $DK$ channels are open; at the physical mass the state lies below the $DK$ threshold and only the isospin-violating $D_s\pi^0$ mode survives, further suppressed by $\epsilon^2$.}
\end{table}

The pattern of Tab.~\ref{tab:abswidths} carries considerable structural information, which we analyze state by state together with the branching fraction ratios of Tab.~\ref{tab:ratiosD1}-\ref{tab:ratiosDs2}.

\emph{The tensor doublet works quantitatively.}\\
The mutual cross predictions $D_2^*(2460)\leftrightarrow D_{s2}^*(2573)$ agree with experiment within $12-13\%$, i.e. within the expected size of SU(3) breaking, the cleanest global validation of the HQET plus wave function framework. Our branching ratio $\Gamma(D_2^*(2460)\to D\pi)/\Gamma(D_2^*(2460)\to D^*\pi) = 2.34$ compares with the measured $1.52\pm0.14$~\cite{PDG2024} and with $2.26$ in Ref.~\cite{Wang:2013tka}.
The residual excess is a known feature of leading order HQET, reduced by $1/m_c$ corrections and by the $D$ wave/$S$ wave interference in the $D^*\pi$ channel.

\emph{Narrow axials, isospin violation and mixing.}\\
The pure $s_\ell = \frac32$ prediction for the $D_1(2420)^0$ (13~MeV) underestimates the measured $31.3\pm1.9$~MeV, while the pure $s_\ell = \tfrac12$ configuration would be broad ($\sim$300~MeV with $g_S$ above). The observed width is reproduced with a mixing angle of a few degrees between the two $1^+$ configurations, as expected from the heavy quark expansion.
The $c\bar{s}$ partner $D_{s1}(2536)^\pm$ is a qualitatively different case. At our model mass of 2498~MeV the state lies marginally below the $D^{*0}K^\pm$ threshold of 2500.5~MeV, so the only channel available in the calculation is the isospin-violating $D_s^{*\pm}\pi^0$ mode. With the suppression factor of Eq.~\eqref{eq:isospin} this yields $1.9\times10^{-3}$~MeV, more than two orders of magnitude below the measured $0.92\pm0.05$~MeV, whereas the naive isospin-symmetric treatment gave $13$~MeV, an order of magnitude above it. The measured width is therefore controlled almost entirely by the isospin conserving $D^{*}K$ $S$ wave, which opens by only $\simeq35$~MeV at the physical mass ($p_f \simeq 0.17$~GeV) and is further reduced by the well known near complete cancellation in that amplitude at the physical $1^+$ mixing angle. Together with the $D_1(2420)^0$, this state is thus a precision probe of the $1^+$ mixing angle and of the position of the $D^*K$ threshold, rather than of the couplings.
The same mechanism resolves what was previously a tension for the $D_{s1}(2460)^\pm$. Its single open channel $D_s^{*\pm}\pi^0$ predicts $1.2\times10^{-3}$~MeV, well within the experimental bound $\leq3.5$~MeV, whereas without $\epsilon^2$, the prediction of $8$~MeV violated that bound by more than a factor of two.

\emph{The $\tfrac12^+$ $D_s$ doublet is incompatible with a conventional $c\bar{s}$ state.}\\
Using the coupling extracted from the broad $D_0^*(2300)^0$, the non-strange member of the same doublet, a conventional $1^3P_0$ $c\bar{s}$ state at present model mass of 2416~MeV would have $\Gamma \sim 335$~MeV, almost two orders of magnitude above the experimental bound $\Gamma(D_{s0}^*(2317)^\pm) \leq 3.8$~MeV. This width is carried entirely by the isospin conserving $D^{0}K^\pm$ and $D^{\pm}K^0$ channels. The isospin-violating $D_s\pi^0$ mode contributes only $1.4\times10^{-2}$~MeV. At the physical mass of 2317.8~MeV the $DK$ channels close and this $\epsilon^2$-suppressed mode is all that remains, which is precisely why the physical state is narrow. The conventional quark model therefore fails as it places the state $\simeq 100$~MeV too high (Sec.~\ref{sec:massanalysis}), and at that mass it predicts a width incompatible with observation by two orders of magnitude. The conjunction of these two statements within a single consistent framework constitutes a quantitative formulation of the $D_{s0}^*(2317)$ puzzle, and supports its interpretation as a state with a dominant $DK$ molecular component.

\emph{$2S$ and $1D$ vectors.}\\
The $D_1^*(2600)^0$ cross prediction (147~MeV vs.\ $93\pm14$~MeV) now overshoots, reflecting the upward shift of $g_H[D_{s1}^*(2700)]$ once the isospin-violating channels are removed from its sum, and its predicted ratio $\Gamma(D^+\pi^-)/\Gamma(D^{*+}\pi^-) = 0.89$ (Tab.~\ref{tab:ratiosD1}) can be compared with the BaBar measurement $0.32\pm0.11$~\cite{BaBar:2010zpy}. Both discrepancies, common to all pure-$2^3S_1$ analyses~\cite{Wang:2013tka}, are resolved by the $2^3S_1-1^3D_1$ mixing, which simultaneously improves the $D_0(2550)$ width noted above, a two parameter explanation of both $S$ wave anomalies. For the $1^3D_1$ $D_1^*(2760)^0$, the large predicted ratio $\Gamma(D\pi)/\Gamma(D^*\pi) = 3.70$ explains naturally why this state was discovered in $D^+\pi^-$, and the $D_{s1}^*(2860)$ cross prediction (175~MeV) is in good agreement with the measured $159\pm80$~MeV.

\emph{Spin-3 states.}\\
The $D_3^*(2750)\to D_{s3}^*(2860)$ prediction (52 vs.\ $53\pm10$~MeV) is excellent, and is a direct consequence of removing the isospin forbidden strength from the $D_{s3}^*(2860)$ channel sum. Our ratio $\Gamma(D\pi)/\Gamma(D^*\pi) = 2.03$ for the $D_3^*(2750)$ agrees with 1.95~\cite{Wang:2013tka} and 1.74~\cite{LHCb:2013jjb}. The analogous $D_{s3}^*(2860)$ ratio $\Gamma(DK)/\Gamma(D^*K) = 1.95$ is a prediction from this study for the ongoing LHCb amplitude analyses.

\emph{The 3-GeV frontier.}\\
For the $D_J(3000)^0$, the measured width ($110.5\pm11.5$~MeV) is reproduced both by the $1F_3$ assignment with $g_Z$ fixed from the $D_2^*(3000)^0$ (99~MeV), and given the $g_H$ spread, by the $3^1S_0$ assignment. The decisive discriminators are the ratios of Tab.~\ref{tab:ratiosD3}, in particular $\Gamma(D_s^{*+}K^-)/\Gamma(D^{*+}\pi^-)$, which is $0.23$ for $1F_3$, $0.08$ for $1F_3'$, and $0.44$ for $3^1S_0$. Combined with the mass spectrum (the $3^1S_0$ level at 3048~MeV matches the observed mass, while the $1F_3$ level lies at 3010~MeV) and the Regge systematics of Sec.~\ref{sec:regge}, we favor the $3^1S_0$ interpretation, with $1F_3$ as the leading alternative. For the natural parity partner $D_J^*(3000)^0$, the $3^3S_1$ and $2^3P_2$ hypotheses both accommodate the measured width for couplings in the extracted range, whereas the $1^3F_4$ assignment would demand $\Gamma \simeq 243$~MeV for $g_R$ fixed from the $D_J(3000)^0$ and predicts a much harder $D\pi$ spectrum. Precise measurements of $\Gamma(D^+\pi^-)/\Gamma(D^{*+}\pi^-)$ ($0.62$ for $3^3S_1$ vs.\ $1.05$ for $2^3P_2$ vs.\ $1.92$ for $1^3F_4$, Tab,~\ref{tab:ratiosD3}) is expected to settle the issue.

\emph{The $D_{s1}^*(2700)^\pm$.}\\
Six of the eight computed channels are open and comparable, the two $D_s^{(*)}\pi^0$ modes being isospin suppressed to the $10^{-4}$ level, and the extracted $g_H = 0.260\pm0.011$ is our most precise $S$-wave coupling. The predicted ratio $\Gamma(DK)/\Gamma(D^*K) \simeq 1.04$ may be compared with the BaBar measurement $\mathcal{B}(D^*K)/\mathcal{B}(DK) = 0.91\pm0.13\pm0.12$~\cite{BaBar:2009rro}, i.e., $\Gamma(DK)/\Gamma(D^*K) = 1.10\pm0.19$. Agreement well within $1\sigma$ strongly supports the $2^3S_1$ assignment over the $1^3D_1$ alternative, for which the ratio would be $4.7$, as in Tab.~\ref{tab:ratiosDs1}. Since this ratio involves only isospin conserving channels, it is unaffected by $\epsilon^2$ and remains the cleanest discriminator available for this state.

\section{Regge trajectories}
\label{sec:regge}

\begin{figure*}[htbp]
\includegraphics[width=0.48\textwidth]{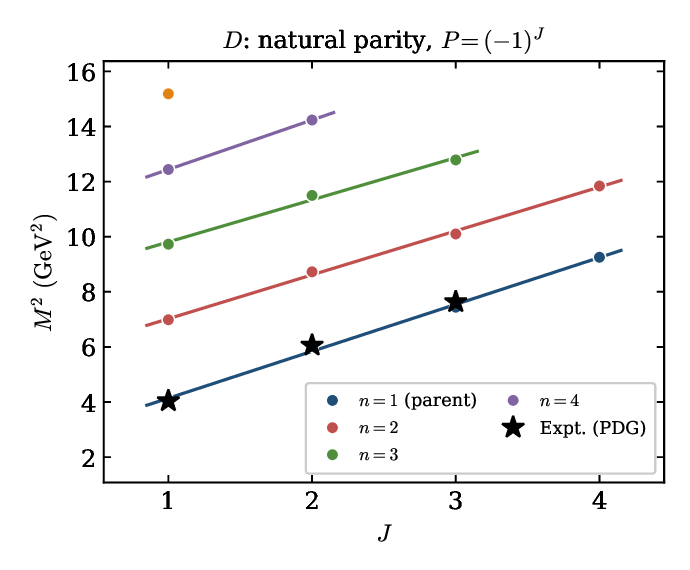}\hfill
\includegraphics[width=0.48\textwidth]{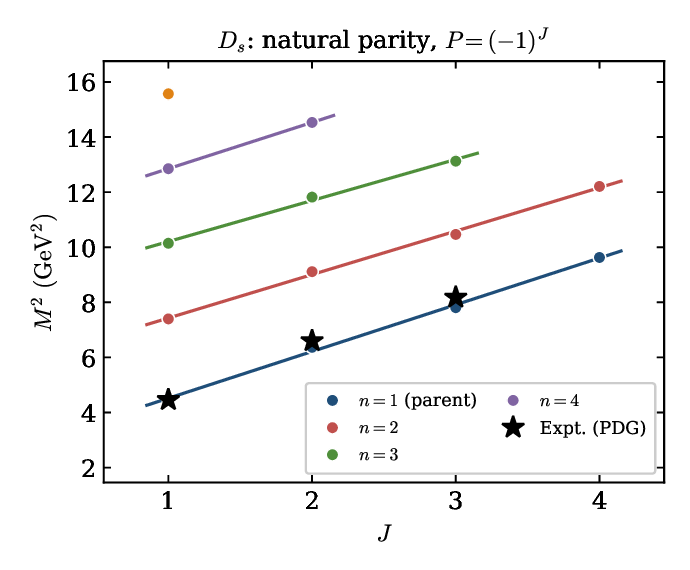}
\caption{Parent and daughter Regge trajectories of the $D$ meson (left) and $D_s$ meson (right) in the $(J,M^2)$ plane for natural-parity states, $P=(-1)^J$.}
\label{fig:reggeJ_nat}
\end{figure*}

\begin{figure*}[htbp]
\includegraphics[width=0.48\textwidth]{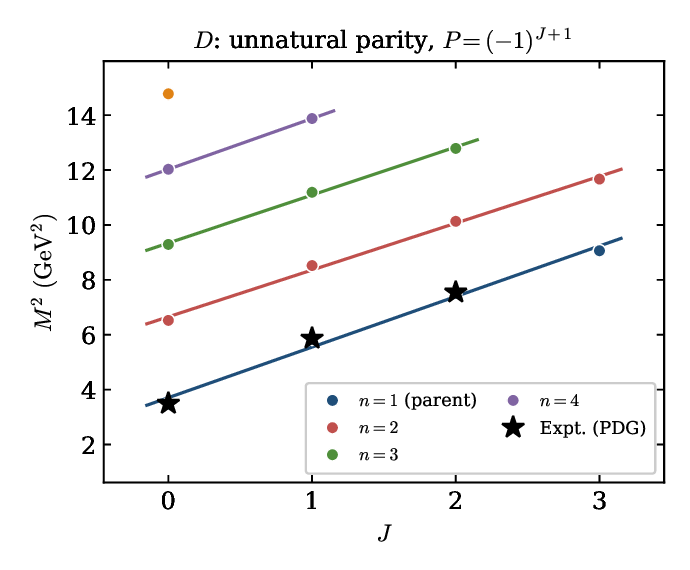}\hfill
\includegraphics[width=0.48\textwidth]{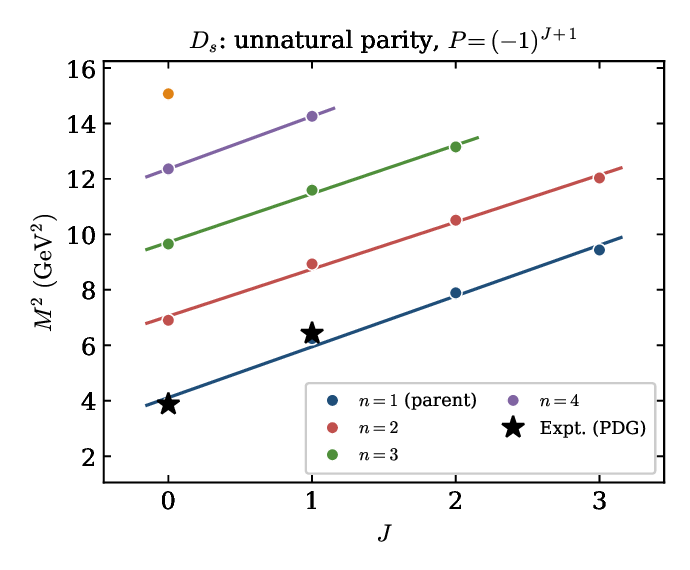}
\caption{Parent and daughter Regge trajectories of the $D$ meson (left) and $D_s$ meson (right) in the $(J,M^2)$ plane for unnatural-parity states, $P=(-1)^{J+1}$.}
\label{fig:reggeJ_unnat}
\end{figure*}

\begin{figure*}[htbp]
\includegraphics[width=0.48\textwidth]{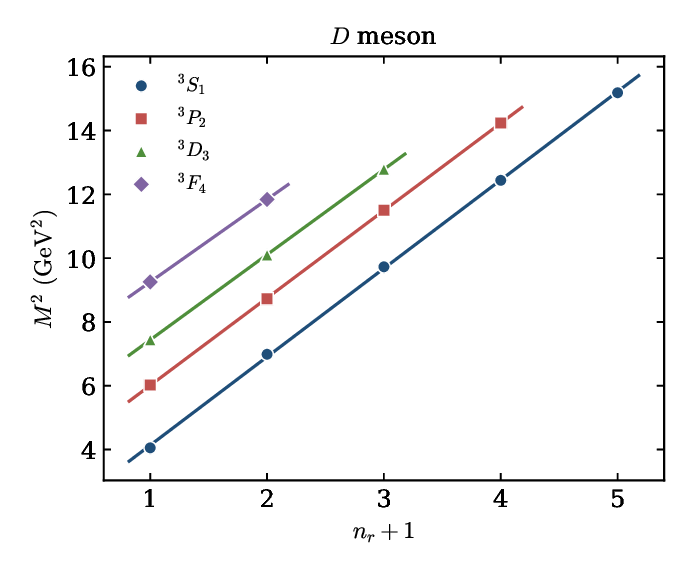}\hfill
\includegraphics[width=0.48\textwidth]{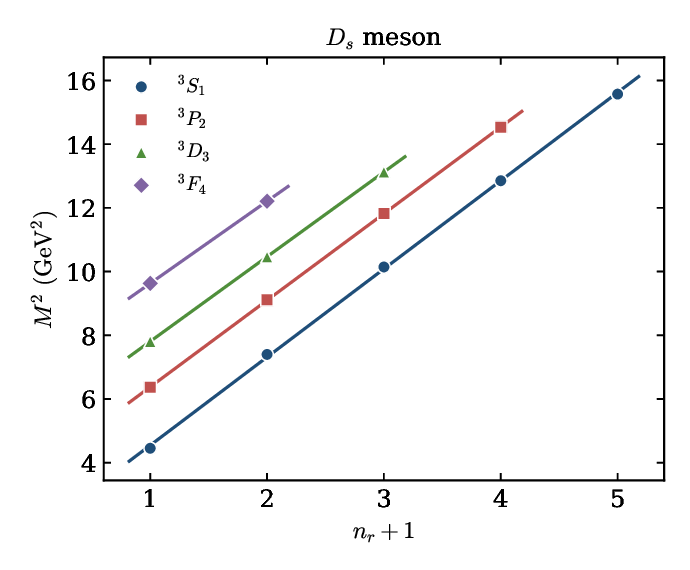}
\caption{Regge trajectories of the $D$ meson (left) and $D_s$ meson (right) in the $(n_r,M^2)$ plane for the $^3S_1$, $^3P_2$, $^3D_3$, and $^3F_4$ states.}
\label{fig:reggeN}
\end{figure*}

Regge trajectories relate the total angular momentum $J$ (or the radial quantum number $n_r = n-1$) of a family of states to the square of their masses and provide a consistency check of the computed spectra as well as a tool for assigning quantum numbers to newly observed states. Using our computed masses, we construct the trajectories~\cite{Ebert:2011jc}
\begin{align}
J &= \alpha\, M^2 + \alpha_0 ,
\label{eq:reggeJ}\\
n_r &= \beta\, M^2 + \beta_0 ,
\label{eq:reggeN}
\end{align}
where $\alpha,\beta$ are the slopes and $\alpha_0,\beta_0$ are the intercepts. Trajectories in the $(J,M^2)$ plane are plotted for the natural parity series $P = (-1)^J$ ($^3S_1$, $^3P_2$, $^3D_3$, $^3F_4$) in Fig.~\ref{fig:reggeJ_nat} and for the unnatural parity series $P = (-1)^{J+1}$ ($^1S_0$, $^1P_1$, $^1D_2$, $^1F_3$) in Fig.~\ref{fig:reggeJ_unnat}.
Trajectories in the $(n_r,M^2)$ plane for the $^3S_1$, $^3P_2$, $^3D_3$, and $^3F_4$ states are shown in Fig.~\ref{fig:reggeN}.

The trajectories, shown together with linear fits and with the measured masses of the established states superposed, are found to be linear and mutually parallel for both the $D$ and $D_s$ families. The fitted slopes and intercepts are given in Tab.~\ref{tab:reggefits}. In the $(J,M^2)$ plane, the parent trajectory slopes are $\alpha = 0.587$ (for $D$ meson) and $0.589$~GeV$^{-2}$ (for $D_s$ meson) for natural parity and $0.543$/$0.546$~GeV$^{-2}$ for unnatural parity. The daughter trajectories have slightly increased slope (by $\lesssim 12\%$) with radial excitation. In the $(n_r,M^2)$ plane the slopes cluster tightly, $\beta = 0.361-0.387$~GeV$^{-2}$ ($D$) and $0.361-0.388$~GeV$^{-2}$ ($D_s$), growing weakly and monotonically with the orbital excitation of the series. Several observations follow.
\begin{itemize}
\item \emph{Magnitude of the slopes:} The universal Regge slope for the light mesons is $\alpha' \simeq 0.88$~GeV$^{-2}$ \cite{Burakovsky:1998zk}. We obtain the slope to be $\simeq 0.55 - 0.65$~GeV$^{-2}$ reflecting the finite mass correction associated with a heavy-light system, in which the heavy quark remains nearly static while the light quark and the confining interaction  contribute to most of the orbital angular momentum. Such nature is also observed by heavy-light Regge systematics using the relativistic quasipotential model \cite{Ebert:2011jc} and with the holographic soft wall studies of charmed mesons \cite{MartinContreras:2020cyg}.

\item \emph{Parallelism and flavor independence:} It is observed that the parent and daughter trajectories are largely parallel, and remarkably the $(n_r,M^2)$ slopes of the $D$ and $D_s$ families coincide almost exactly series by series. This is a direct imprint of the flavor independence of the confining interaction. It is also found that the strange quark mass shifts the intercepts ($\beta_0$ differs by $\simeq 0.15$ between the two families, corresponding to the $\sim100$~MeV mass difference) but not the slopes as expected, if the linear confining term $Ar$ dominates the excitation dynamics.

\item \emph{Departures from linearity:} The only visible curvature occurs at the low mass ends of the $^3D_3$, $^3P_2$, $^3D_1$, and $^3F_4$ series where the states are most affected by the fine structure terms. This observation is consistent with the general expectation that trajectories of mesons containing strange and/or heavy quarks are nonlinear at small $M^2$ and asymptotically linear~\cite{MartinContreras:2020cyg}.

\item \emph{Experimental overlay and assignments:} The experimental measured masses of the established states shown by stars in Figs.~\ref{fig:reggeJ_nat} and \ref{fig:reggeJ_unnat} fall on our parent trajectories within the acceptable deviation range, except for the $D_3^*(2750)$/$D_{s3}^*(2860)$. Extrapolating along the fitted lines, the $M^2$ values of the $D_J(3000)^0$, $D_J^*(3000)^0$, and $D_2^*(3000)^0$ land on the $n=3$ unnatural $S$, $n=3$ natural $S$, and $n=3$ $^3P_2$ trajectories respectively. This assignment argument is independent of the decay analysis of Sec.~\ref{sec:strong}. The same fits predict the fourth radial excitations $4^1S_0$ and $4^3S_1$ at $\simeq3.47$ and $3.53$~GeV as shown in Tab.~\ref{tab:massD_SP}, which are directly testable in future high statistics $D^{(*)}\pi$ amplitude analyses.
\end{itemize}

\begin{table}[htbp]
\caption{Fitted Regge slopes and intercepts for the parent trajectories, Eqs.~\eqref{eq:reggeJ} and \eqref{eq:reggeN}. Nat.\ (unnat.)\ denotes the natural (unnatural) parity series in the $(J,M^2)$ plane; the $(n_r,M^2)$ fits are per orbital series. \label{tab:reggefits}}
\begin{tabular*}{\textwidth}{@{\extracolsep{\fill}}llcc@{}}
\hline\hline
Plane / series & Family & Slope (GeV$^{-2}$) & Intercept \\
\hline
$(J,M^2)$, nat., $n=1$ & $D$ & $\alpha=0.587$ & $\alpha_0=-1.431$ \\
 & $D_s$ & $\alpha=0.589$ & $\alpha_0=-1.662$ \\
$(J,M^2)$, unnat., $n=1$ & $D$ & $\alpha=0.543$ & $\alpha_0=-2.014$ \\
 & $D_s$ & $\alpha=0.546$ & $\alpha_0=-2.243$ \\
$(n_r,M^2)$, $^3S_1$ & $D$ & $\beta=0.361$ & $\beta_0=-1.490$ \\
 & $D_s$ & $\beta=0.361$ & $\beta_0=-1.640$ \\
$(n_r,M^2)$, $^3P_2$ & $D$ & $\beta=0.365$ & $\beta_0=-2.192$ \\
 & $D_s$ & $\beta=0.368$ & $\beta_0=-2.344$ \\
$(n_r,M^2)$, $^3D_3$ & $D$ & $\beta=0.374$ & $\beta_0=-2.782$ \\
 & $D_s$ & $\beta=0.376$ & $\beta_0=-2.935$ \\
$(n_r,M^2)$, $^3F_4$ & $D$ & $\beta=0.387$ & $\beta_0=-3.577$ \\
 & $D_s$ & $\beta=0.388$ & $\beta_0=-3.733$ \\
 \hline\hline
\end{tabular*}
\end{table}

\section{Results and discussion}
\label{sec:results}
Having determined the confinement strengths, quark masses ($m_{u/d}, ~m_s$) and taking the inputs of charm quark masses from our previous work \cite{Soni:2017wvy}, we compute the mass spectra of $D$ and $D_s$ mesons considering the interaction of the Cornell potential form. The computed spectra are listed in Tab.~\ref{tab:massD_SP} - \ref{tab:massDs_DF}. In Sec.~\ref{sec:massanalysis}, we have made brief analysis of our computed spectra in comparison with different theoretical predictions and available experimental data. Moreover, in Tab.~\ref{tab:devD} and \ref{tab:devDs}, we list experimental established states and compare our predictions and also provide the \% deviation. It is observed that the thirteen tabulated $D$-meson states are reproduced with deviations between $0.02\%$ and $1.3\%$, with a mean of $0.5\%$. Similar is the observation for $D_s$ mesons with the deviation between 0.03\% to 4.28\% with a mean of 1.4\% deviation.
It is worth noting here that we fit only four $1S$ ground masses to determine the model parameters and all the excited levels of $D$ and $D_s$ are free predictions.

With the help of computed masses, model parameters, and numerical wavefunctions, we compute the leptonic decay constants in Sec.~\ref{sec:decayconst} and leptonic decay widths in Sec.~\ref{sec:leptonic}. We present our predictions in Tab.~\ref{tab:fpD}-\ref{tab:leptonicBF}.
It is observed that our result for $f_D$ is within 9\% of the experimental data and for $f_{D_s}$, our prediction is in excellent agreement with the PDG data. Similarly, our predictions for the leptonic branching fractions are in very good agreement with the experimental data for $D_s$ meson and for $D$ meson, our results overestimate the experimental data.
In Sec~\ref{sec:em}, we study the electromagnetic decay widths and our results are tabulated in Tab.~\ref{tab:E1D} - \ref{tab:M1Ds} and compare with the different approaches. It is observed that our predictions are in line with the data reported in literature.
We further compute the strong decay widths of $D$ and $D_s$ mesons in the heavy quark effective theory framework in Sec.~\ref{sec:strong}. In the charm-strange sector all $D_s^{(*)}\pi^0$ transitions are isospin violating, and we consistently weight them by the suppression factor $\epsilon^2 \simeq 1.5\times10^{-4}$ of Eq.~\eqref{eq:isospin}, which is the charm analogue of the mechanism operating in the strange-bottom scalar sector. Our results of strong decay widths in the units of square coupling and ratios of the partial widths are listed Tab.~\ref{tab:strongD1} - \ref{tab:strongDs2}. We further extract the heavy quark effective coupling and compute the absolute coupling in Sec.~\ref{sec:couplings} and results are listed in Tab.~\ref{tab:couplings} and \ref{tab:abswidths}.
We also plot the Regge trajectories using our computed masses and provide the slopes and intercepts in Tab. \ref{tab:reggefits}.

The present study not only reproduces the observed masses and decay properties, but also it helps in identifying the nature of the excited charmed mesons.
It suggests that the $D_2(2740)^0$ is a mixed $2^-$ state and determines the mixing angles of the $1^+$ states using the measured widths of the $D_1(2420) $ and $D_{s1}(2536)$.
The model also shows that the $D_{s0}(2317)$ cannot be explained as a simple $c\bar{s}$ meson because its predicted mass is about 100 MeV higher than the observed value and, at that mass, its predicted decay width of $\sim335$~MeV exceeds the experimental limit by almost two orders of magnitude.
Similarly, the measured $DK/D^*K$ branching ratio supports assigning the $D_{s1}^*(2700)$ as the $2^3S_1$ state rather than a $D$ state wave.
By combining the information from the mass spectrum, decay widths, and Regge trajectories, our study further identifies the three observed states near 3 GeV as the $3^1S_0, 3^3S_1$ and $3^3P_2$ states.

The spectrum alone could be fit by an inadequate wave function; the decay properties exclude this. The same wave functions yield $f_{D_s}$ within $0.4\%$ of experiment and $f_D$ within $9\%$ (Sec.~\ref{sec:fanalysis}); leptonic branching fractions correct at the $1-25\%$ level with the deviations quantitatively traced to $f_{D_{(s)}}^2$ (Sec.~\ref{sec:leptanalysis}); radiative widths within the computated results (Sec.~\ref{sec:emanalysis}); and most stringently strong width coefficients that, once the HQET couplings are extracted, cross predict total widths between the $D$ and $D_s$ sectors at the $2-12\%$ level for the tensor, $D$-wave and $F$-wave doublets and pass the heavy quark flavor symmetry test coupling by coupling as shown in Tab.~\ref{tab:couplings} and \ref{tab:abswidths}. The $2S$ doublet, where the three independent $g_H$ determinations still span a factor of 1.5, is an outlier and is discussed in Sec.~\ref{sec:couplings}.

The framework identifies the $D_2(2740)^0$ as a mixed $2^-$ state, quantifies the $1^+$ mixing angles through the $D_1(2420)$/$D_{s1}(2536)$ widths, converts the $D_{s0}^*(2317)$ anomaly into a statement relating a $100$~MeV mass deficit and a two order of magnitude width conflict under the $c\bar{s}$ hypothesis. This resolves the $D$-wave/$S$-wave competition for the $D_{s1}^*(2700)$ in favour of $2^3S_1$ via the measured $DK/D^*K$ ratio, and through the concordance of masses, width patterns, and Regge systematics assigns the three structures near 3~GeV to $3^1S_0$, $3^3S_1$, and $3^3P_2$.
The model deviates from the data in case of $^3D_3$ masses, the $f_{D^*}/f_D$ hierarchy, and the $D_0(2550)$ width. While the disagreements are systematic, they point to specific $j$-dependent spin-orbit strength, wave function level hyperfine coupling as well as $2S-1D$ mixing.

\subsection{Model limitations and uncertainties}
\label{sec:uncertainties}

Several sources of theoretical uncertainty should be kept in mind when using these predictions. (i)~The nonrelativistic approximation, while validated at the level of the wave functions against relativistic treatments~\cite{Godfrey:2015dva,Ebert:2009ua}, becomes less reliable for the higher excitations, where the light quark is increasingly relativistic. (ii)~The Gaussian smearing parameters $\sigma_D$ and $\sigma_{D_s}$ are fitted to the hyperfine splittings of the ground states. Varying them within the range that keeps the $1S$ hyperfine splitting within its experimental uncertainty changes the excited state masses by a few MeV and the decay constants slightly. (iii)~The $E1$/$M1$ widths are computed in the leading nonrelativistic multipole approximation and carry the usual model dependence of the wave functions. (iv)~The strong widths rely on the leading order HQET Lagrangians. Hence, $1/m_c$ corrections may be sizable for individual channels, though they largely cancel in the ratios. (v)~The isospin-breaking factor of Eq.~\eqref{eq:isospin} is a leading order estimate based on $\pi^0-\eta$ mixing at the quark mass level. It carries an uncertainty of about $30\%$ from the current quark mass ratios and from higher order chiral corrections, which propagates directly into the $D_s^{(*)}\pi^0$ partial widths but has a negligible effect on the extracted couplings as they are dominated by isospin conserving channels. Finally, the $j^P = \frac12^+$ $D_{s0}^*(2317)-D_{s1}(2460)$ doublet is poorly described by any pure $c\bar{s}$ picture, our results for these states should be interpreted accordingly.

\section{Conclusions and outlook}
\label{sec:conclusion}

We have presented a unified and self consistent study of the $D$ and $D_s$ meson families within a Cornell potential model including a Gaussian smeared hyperfine interaction term. Using a single set of wave functions, we have investigated their spectroscopy, decay constants, leptonic decays, electromagnetic transitions, strong decays, and Regge trajectories in a consistent manner. The main conclusions of the present work are summarized as follows:
\begin{enumerate}
\item With seven parameters fixed to ground state data alone, the model reproduces every confirmed $D$ and $D_s$ level with a mean deviation of $0.5\%$ and $0.7\%$, respectively (excluding the exotic candidate $\frac12^+$ $D_s$ doublet), and predicts the presently unobserved $3S-5S$, $2P-4P$, $2D-3D$, and $1F-2F$ levels as concrete search targets for the charm factories.
\item All isospin-violating $D_s^{(*)}\pi^0$ channels have been weighted by the suppression factor $\epsilon^2 = \frac{3}{16}\left[(m_d-m_u)/(m_s-\frac{m_u+m_d}{2})\right]^2 \simeq 1.5\times10^{-4}$. Imposing this factor removes isospin forbidden strength from the $D_s$ channel sums and thereby sharpens the flavor symmetry test appreciably. The $D$ and $D_s$-sector determinations of $g_X$ and $g_Y$, which previously differed by $13\%$ and $15\%$, now agree to $4\%$ and $1\%$ respectively, and the $D_{s3}^*(2860)$ and $D_{s1}(2460)$ widths are brought into agreement with experiment.
\item The full set of HQET strong couplings, $g_H = 0.21-0.32$, $g_S = 0.61$, $g_T = 0.39-0.42$, $g_X = 0.21-0.22$, $g_Y = 0.39-0.40$, $g_Z = 0.26$, and $g_R = 0.20$, has been extracted from the measured total widths. The independent $D$ and $D_s$ sector extractions agree coupling by coupling for the $P$, $D$ and $F$-wave doublets, providing a quantitative confirmation of heavy quark flavor symmetry in the charm sector. This results into absolute width predictions among the two families at the $2-12\%$ level, while only the $2S$ doublet remains discrepant suggesting $2S-1D$ mixing.
\item The analysis sharpens several structural conclusions. The $D_{s0}^*(2317)^\pm$ is incompatible with a conventional $c\bar{s}$ interpretation by both mass ($\sim100$~MeV) and width (two orders of magnitude). The $D_1(2420)$, $D_{s1}(2536)$, and $D_2(2740)$ widths measure the $1^+$ and $2^-$ mixing angles, the $D_{s1}(2536)$ additionally probes the position of the $D^*K$ threshold once its isospin-violating mode is properly suppressed. The $D_{s1}^*(2700)^\pm$ is confirmed as the $2^3S_1$ state by the measured $DK/D^*K$ ratio. The concordance of the mass spectrum, the width patterns, and the linear, parallel Regge trajectories with slopes $\alpha \simeq 0.55-0.65$~GeV$^{-2}$, $\beta \simeq 0.36-0.39$~GeV$^{-2}$, assigns the $D_J(3000)^0$, $D_J^*(3000)^0$, and $D_2^*(3000)^0$ to the $3^1S_0$, $3^3S_1$, and $3^3P_2$ states, respectively.
\end{enumerate}

The importance of these results extends beyond spectroscopy. The validated wave functions and decay constants feed directly into semileptonic form factor calculations and CKM extractions, the excited state predictions define the $D^{**}$ background model for the $R(D^{(*)})$ program, and the quantified boundary of the conventional $c\bar q$ description tells experiments precisely where exotic dynamics must be invoked.

Looking forward, three experimental measurements would be most decisive: (a)~the ratios $\Gamma(D_s^{*}K)/\Gamma(D^{*}\pi)$ and $\Gamma(D\pi)/\Gamma(D^{*}\pi)$ for the states near 3~GeV, which discriminate sharply among the $3S$, $2P$, and $1F$ hypotheses (Tab.~\ref{tab:ratiosD3}), (b)~any $1P\to1S\gamma$ radiative branching fraction, which would arbitrate the two orders of magnitude spread among model wave functions, and (c)~high precision line shapes of the $D_0(2550)^0$ and $D_1^*(2600)^0$, which test the $2S-1D$ mixing scenario that resolves the $g_H$ tension found here.

On the theory side, natural extensions of this work are the inclusion of $1/m_c$ corrections and channel couplings for the $\frac12^+$ doublets, a wave function level treatment of the hyperfine interaction to correct the $f_{V}/f_{P}$ hierarchy, and the transfer of the identical framework with no new parameters beyond $m_b$ to the $B$ and $B_s$ families, where LHCb and Belle~II data are accumulating rapidly and where the same isospin-breaking factor governs the $B_s(1^3P_0)\to B_s\pi$ transition.

\section*{Acknowledgement}
JNP acknowledges the high performance computing cluster facility at Sardar Patel University, Vallabh Vidyanagar, procured under the PM-USHA Scheme of the Government of India.

\bibliography{apssamp}
\end{document}